\documentstyle[preprint,aps,graphicx,floats]{revtex}

\begin{document}

\preprint{$
\begin{array}{l}
\mbox{UB-HET-09-02}\\[-3mm]
\mbox{May~2009} \\ [3mm]
\end{array}
$}

\title{Measuring the Higgs Boson Self-coupling at High Energy $e^+e^-$
Colliders}

\author{U.~Baur\footnote{baur@ubhex.physics.buffalo.edu}}
\address{Department of Physics,
State University of New York, \\
Buffalo, NY 14260, USA}

\maketitle 

\begin{abstract}
\baselineskip13.pt  
Standard Model Higgs pair production at $e^+e^-$ colliders has the
capability to determine the Higgs boson self-coupling $\lambda$. I
present a detailed analysis of the $e^+e^-\to ZHH$ and
$e^+e^-\to\nu\bar\nu HH$ signal channels, and the relevant background
processes, for future $e^+e^-$ linear colliders with center of mass
energies of $\sqrt{s}=0.5$~TeV, 1~TeV, and 3~TeV. Special attention is
given to the role non-resonant Feynman diagrams play, and the
theoretical uncertainties of signal and background cross sections. I
also derive 
quantitative sensitivity limits for $\lambda$. I find that an $e^+e^-$
collider with $\sqrt{s}=0.5$~TeV can place meaningful bounds on
$\lambda$ only if the Higgs boson mass is relatively close to its
current lower limit. At an $e^+e^-$ collider with $\sqrt{s}=1$~TeV
(3~TeV), $\lambda$ can be determined with a precision of $20-80\%$
($10-20\%$) for integrated luminosities in the few ab$^{-1}$ range and
Higgs boson masses in the range $m_H=120-180$~GeV.

\end{abstract}

\newpage


\tightenlines

\section{Introduction}
\label{sec:one}
The CERN Large Hadron Collider (LHC) is scheduled to begin operation
in 2009, beginning a new era wherein the mechanism of electroweak
symmetry breaking and fermion mass generation will be revealed and
studied in detail.  Although alternative mechanisms exist in
theory, this is generally believed to be a light Higgs boson with mass
$114~{\rm GeV}<m_H<145$~GeV~\cite{ewk08,Phenomena:2009pt,Goebel:2009qy}.
More specifically, we 
expect a fundamental scalar sector which undergoes spontaneous symmetry
breaking as the result of a potential which acquires a nonzero vacuum
expectation value.  The LHC will easily find a light Standard Model
(SM) Higgs boson with moderate luminosity~\cite{wbf_ww,wbf_exp}.
Moreover, the LHC will have the capability to determine some
of its properties~\cite{atlas_tdr,cms_tdr}, such as its fermionic and
bosonic decay modes and couplings~\cite{wbf_ll,Hcoup,Yb,Yt}, including
invisible decays~\cite{wbf_inv} and possibly even rare decays to
second generation fermions~\cite{Hmumu}. An $e^+e^-$ linear
collider with a center of mass energy of 350~GeV or more will be able to
significantly improve these preliminary measurements, in some cases
by an order of magnitude in precision, if an integrated luminosity
of 500~fb$^{-1}$ can be achieved~\cite{LC}.

Perhaps the most important measurement after a Higgs
boson discovery is of the Higgs potential itself, which requires
measurement of the trilinear and quartic Higgs boson self-couplings.
Only multiple Higgs boson production can probe these
directly~\cite{higgs_self,LC_HH1a,LC_HH1}. Several years ago, studies
exploring the potential of 
the LHC, a luminosity-upgraded LHC (SLHC) with roughly
ten times the amount of data expected in the first run, and a
Very Large Hadron Collider (VLHC), have appeared in the
literature~\cite{SLHC,BPR,BPR2,Baur:2003gpa,blondel}. There are also
numerous quantitative sensitivity limit analyses of Higgs 
boson pair production in $e^+e^-$ collisions ranging from 500~GeV to
3~TeV center of mass
energies~\cite{LC_HH1a,LC_HH1,LC_HH2a,LC_HH2,LC_HH3,LC_HH4,Yasui:2002se,Boumediene:2008hs,Takubo:2009rn,Giannelli:2009fe},
and for $\gamma\gamma\to HH$~\cite{Belusevic:2004pz}.
The $e^+e^-$ studies, usually, focus on one particular Higgs mass and/or final
state, only one center of mass energy, and, in many cases, estimate the
background using a leading-log shower approximation. Furthermore, the
effects of non-resonant Feynman diagrams are not taken into account.

In this paper, I present a more thorough investigation of
Higgs boson pair production in $e^+e^-$ collisions. I calculate the
$e^+e^-\to ZHH$ and $e^+e^-\to\nu\bar\nu HH$
($\nu=\nu_e,\,\nu_\mu,\,\nu_\tau$) signal cross sections for
$\sqrt{s}=0.5$~TeV, 1~TeV and 3~TeV, and $m_H=120$~GeV, 140~GeV and
180~GeV, although $m_H=180$~GeV is disfavored by the most recent fit to
electroweak data when direct search limits from the Tevatron experiments
are taken into account~\cite{Goebel:2009qy}. A center of mass energy of
$0.5-1$~TeV is considered for the 
International Linear Collider (ILC)~\cite{Brau:2007zza}, whereas
$\sqrt{s}=3$~TeV is the target energy for CERN's CLIC
concept~\cite{Braun:2008zz}. Only unpolarized electron and positron
beams are considered. Since the cross section for the one-loop process
$e^+e^-\to HH$ is more than one order of magnitude smaller than that for
$ZHH$ production~\cite{LopezVillarejo:2008xw}, it is not considered
here. Likewise, I will ignore the process $e^+e^-\to e^+e^-HH$; due to
the small $Zee$ coupling, its
cross section is about a factor~6 less than that for $\nu\bar\nu HH$
production~\cite{GutierrezRodriguez:2008nk}. 
I consider several final states, and estimate both the reducible and
irreducible backgrounds using exact matrix element calculations. Special
attention is given to the role non-resonant Feynman diagrams play, and
the theoretical uncertainties of signal and background cross sections. All
calculations are performed taking into account the anticipated resolution
of future $e^+e^-$ detectors. Finally, I derive quantitative sensitivity
bounds for several integrated luminosities, and compare the capabilities
of the considered $e^+e^-$ colliders with each other and those of the
LHC, a luminosity upgraded LHC (SLHC)~\cite{SLHC}, and a Very Large
Hadron Collider operating at a center of mass energy of
200~TeV~\cite{ambrosio}. 

The remainder of this paper is organized as follows. I first review the
definition of the Higgs boson self-couplings and briefly discuss SM and
non-SM predictions for these parameters in Sec.~\ref{sec:two}. The
methods and tools used in my calculations, together with the
parametrized detector resolution are summarized in
Sec.~\ref{sec:three}. In Secs.~\ref{sec:four} and~\ref{sec:five} the
$ZHH$ and $\nu\bar\nu HH$ signal channels and all relevant backgrounds
are discussed. Quantitative sensitivity limits are calculated in
Sec.~\ref{sec:six}. In Sec.~\ref{sec:seven}, I finally present my
conclusions. 

\section{Higgs boson self-couplings}
\label{sec:two}

The trilinear and quartic Higgs boson couplings $\lambda$ and
$\tilde\lambda$ are defined through the potential
\begin{equation}
\label{eq:Hpot}
V(\eta_H) \, = \,
{1\over 2}\,m_H^2\,\eta_H^2\,+\,\lambda\, v\,\eta_H^3\,+\,{1\over 4}\,
\tilde\lambda\,\eta_H^4 ,
\end{equation}
where $\eta_H$ is the physical Higgs field, $v=(\sqrt{2}G_F)^{-1/2}$
is the vacuum expectation value, and $G_F$ is the Fermi constant.  In
the SM the self-couplings are
\begin{equation}
\label{eq:lamsm}
\tilde\lambda=\lambda=\lambda_{SM}={m_H^2\over 2v^2}\,.
\end{equation}
Regarding the SM as an effective theory, the Higgs boson
self-couplings $\lambda$ and $\tilde\lambda$ are {\it per se} free
parameters, and $S$-matrix unitarity constrains $\tilde\lambda$ to
$\tilde\lambda\leq 8\pi/3$~\cite{unit}.  Since future collider
experiments likely cannot probe $\tilde\lambda$~\cite{Plehn:2005nk}, I
concentrate on the 
trilinear coupling $\lambda$ in the following.  The quartic Higgs
coupling does not affect the Higgs pair production processes I
consider.

In the SM, radiative corrections decrease $\lambda$ by $4-11\%$ for
$120~{\rm GeV}< m_H<200$~GeV~\cite{yuan}.  Larger deviations are
possible in scenarios beyond the SM. For example, in two Higgs doublet
models where the lightest Higgs boson is forced to have SM like
couplings to vector bosons, quantum corrections may increase the
trilinear Higgs boson coupling by up to $100\%$~\cite{yuan}.  In the
MSSM, loop corrections modify
the self-coupling of the lightest Higgs boson in the decoupling limit,
which has SM-like 
couplings, by up to $8\%$ for light stop squarks~\cite{hollik}.
Anomalous Higgs boson self-couplings also appear in various other
scenarios beyond the SM, such as models with a composite Higgs
boson~\cite{georgi}, or in Little Higgs models~\cite{lhiggs}.  In
many cases, the anomalous Higgs boson self-couplings can be
parametrized in terms of higher dimensional operators which are
induced by integrating out heavy degrees of freedom.  A systematic
analysis of Higgs boson self-couplings in a higher dimensional
operator approach can be found in Ref.~\cite{tao}.

\section{Calculational details}
\label{sec:three}

All calculations presented here have been performed at tree-level using {\tt
MadEvent}~\cite{Maltoni:2002qb} which has been modified to allow for 
non-standard values of $\lambda$. For some background calculations it
was also necessary to increase the maximum number of Feynman diagrams
and/or configurations allowed so that {\tt MadEvent} could be used
successfully. Background 
calculations involving the strong coupling constant, $\alpha_s$, were
performed with the renormalization scale set to $\mu=M_Z$, where $M_Z$
is the mass of the $Z$-boson. The SM
parameters used are~\cite{Mangano:2002ea}
\begin{eqnarray}
\label{eq:input1}
G_{\mu} = 1.16639\times 10^{-5} \; {\rm GeV}^{-2}, & \quad & \\
M_Z = 91.188 \; {\rm GeV}, & \quad & M_W=80.419  \; {\rm GeV}, \\
\label{eq:input2} 
\sin^2\theta_W=1-\left({M^2_W\over M_Z^2}\right), & \quad &
\alpha_{G_\mu} = {\sqrt{2}\over\pi}\,G_F \sin^2\theta_W M_W^2,
\end{eqnarray}
where $G_F$ is the Fermi constant, $M_W$ is the $W$ mass,
$\theta_W$ is the weak mixing angle, and
$\alpha_{G_\mu}$ is the electromagnetic coupling constant in the $G_\mu$
scheme. 

The basic kinematic acceptance cuts for events at the ILC and CLIC are
\begin{eqnarray}
\label{eq:cuts1}
\nonumber &
E_\ell>15~{\rm GeV},\qquad E_j>15~{\rm GeV},\qquad p\llap/_T>15~{\rm GeV,}\\
 &
5^\circ <\theta(j,{\rm beam})<175^\circ\, , \qquad \theta(j,j')>10^\circ
,\\
\nonumber &
5^\circ <\theta(\ell,{\rm beam})<175^\circ\, , \qquad \theta(\ell,j)>10^\circ ,
\end{eqnarray}
where $\theta(k,l)$ is the angle between the two objects $k$ and $l$,
$\ell=e,\,\mu$, $E$ is the energy, and $p\llap/_T$ is the missing
transverse momentum. The $p\llap/_T$ cut is only imposed on final states
where there are one or more neutrinos present. Likewise, the angular and
energy cuts on the charged leptons are only imposed if there is at least
one charged lepton in the event which does not result from the decay of
a charm or bottom quark.

In all calculations, I include minimal detector effects by Gaussian
smearing of the parton momenta according to ILC detector
expectations~\cite{Behnke:2007gj}
\begin{eqnarray}
{\Delta{E} \over E}{\rm (had)} 
&=& {0.405 \over \sqrt{E}} ,
\nonumber \\[2.mm] 
{\Delta{E} \over E}{\rm (lep)} 
&=& {0.102 \over \sqrt{E}},
\end{eqnarray}
and assume that charged leptons and jets can be detected with an
efficiency close to 100\%. For the $b$-tagging efficiency, $\epsilon_b$,
and the
associated misidentification probabilities of light quark/gluon jets and
$c$-quarks 
to be tagged as a $b$-quarks, $P_{j\to b}$ and $P_{c\to b}$, I consider
two scenarios~\cite{Behnke:2001qq,hansen}:
\begin{equation}
\label{eq:eff1}
\epsilon_b=90\%~{\rm with}~P_{j\to b}=0.5\%~{\rm and}~P_{c\to
b}=10\%,
\end{equation}
and
\begin{equation}
\label{eq:eff2}
\epsilon_b=80\%~{\rm with}~P_{j\to b}=0.1\%~{\rm and}~P_{c\to
b}=2\%.
\end{equation}
The energy loss in $b$-jets due to $b$-quark decays is taken into
account via a parametrized 
function~\cite{Baur:2007ck}, and I will assume that $b$- and $\bar
b$-quarks can be distinguished with 100\% efficiency. This should be a
good approximation since ILC detectors are expected to be able to 
measure the electric charge of a $b$-jet with an efficiency of 90\% or
better~\cite{hansen,Hillert:2005rp}. 

The identification of $H\to b\bar b$ and hadronic $W$- and $Z$-decays
plays an important role in separating the $ZHH$ and $\nu\bar\nu HH$
signal from the background. To identify $W\to jj$ and $Z\to jj$ decays,
I require ($V=W,\, Z$)
\begin{equation}
\label{eq:mjj}
|M_V-m(jj)|<8~{\rm GeV,}
\end{equation}
where $m(jj)$ is the di-jet invariant mass. Since the energy loss of the
$b$-jets distorts the $H\to b\bar b$ Breit-Wigner function and lowers
the $b\bar b$ invariant mass, I impose
\begin{eqnarray}
\label{eq:mbb}
100~{\rm GeV}<m(b\bar b)<126~{\rm GeV} & \qquad & {\rm for}~m_H=120~{\rm
GeV~and}, \\ \label{eq:mbb1}
120~{\rm GeV}<m(b\bar b)<150~{\rm GeV} & \qquad & {\rm for}~m_H=140~{\rm
GeV.}
\end{eqnarray}
This captures most of the signal cross section. I also assume that, by
the time Higgs pair 
production is being analyzed, the Higgs boson mass is accurately known
from experiments at the LHC and/or an analysis of the process $e^+e^-\to
ZH$. Initial state radiation and beamstrahlung are not included in any
of the calculations presented here.

In order to derive sensitivity limits for $\lambda$, the distribution of
Higgs-pair invariant mass, $M_{HH}$, will be used. The $M_{HH}$
distribution is known to be sensitive to the Higgs boson self-coupling,
in particular for small values of the Higgs-pair mass~\cite{LC_HH3}.

It is well known~\cite{LC_HH1a,LC_HH2a} that the $e^+e^-\to ZHH$ cross 
section decreases with increasing center of mass energy while 
$\sigma(e^+e^-\to\nu\bar\nu HH)$ increases. For $\sqrt{s}=0.5$~TeV,
$\sigma(e^+e^-\to ZHH)\gg\sigma(e^+e^-\to\nu\bar\nu HH)$, and for
$\sqrt{s}=3$~TeV, $\sigma(e^+e^-\to ZHH)\ll\sigma(e^+e^-\to\nu\bar\nu
HH)$. For $\sqrt{s}=1$~TeV, the two cross sections are of the same
order of magnitude, with $e^+e^-\to\nu\bar\nu
HH$ being the larger source of Higgs boson pairs. I, therefore, consider
$ZHH$ ($\nu\bar\nu HH$) 
production for $\sqrt{s}\leq 1$~TeV ($\sqrt{s}\geq 1$~TeV) only. For a
center of mass energy of 0.5~TeV, Higgs pair production is strongly
phase space suppressed if $m_H>140$~GeV~\cite{Baur:2003gpa}. The
$m_H=180$~GeV case, therefore, is only analyzed for $\sqrt{s}\geq 1$~TeV. 

As stated before, 
all calculations reported here are carried out at tree level. The
complete one-loop radiative corrections to $e^+e^-\to ZHH$ and
$e^+e^-\to\nu\bar\nu HH$ are
known~\cite{Belanger:2003ya,Boudjema:2005rk} to modify the lowest cross
section by a few percent for the energy range considered in this
paper. As I will show, for most of the final states considered in this
paper, there are uncertainties which are significantly larger than the
effect of the one-loop electroweak radiative corrections. Electroweak
radiative corrections thus will be ignored in the following. 

\section{$ZHH$ analysis}
\label{sec:four}

The total $e^+e^-\to ZHH$ cross section at $\sqrt{s}=0.5$~TeV is
$\sigma(ZHH)\approx 0.18$~fb. It decreases with
increasing center of mass energy and Higgs boson mass.
The current design luminosity of the ILC 
is~\cite{Brau:2007zza} ${\cal L}=2\times 10^{34}\,{\rm cm}^{-2}{\rm
s}^{-1}$, corresponding to a yearly integrated luminosity of
200~fb$^{-1}$. One therefore expects a total of about 40~$ZHH$ events
per year at such a machine, before $Z$- and Higgs boson decays, and
experimental acceptance cuts and efficiencies, are taken into account. 

With an expected $b$-tagging efficiency of 80\% or higher, and a fairly large
branching ratio ($B(H\to b\bar b)\approx 68\%$ (33\%) for $m_H=120$~GeV
($m_H=140$~GeV)), requiring $HH\to b\bar bb\bar b$  and $Z\to\ell^+\ell^-$
($\ell=e,\,\mu$) offers the best chance to identify $ZHH$
events. Unfortunately, the $Z\to\ell^+\ell^-$ branching ratio is very
small, and too few $\ell^+\ell^- b\bar bb\bar b$ events are left to make
this final 
state viable for a measurement of the Higgs boson self-coupling. I therefore
concentrate on the $ZHH\to jjb\bar bb\bar b$ final state, where 
$jj$ denotes a light jet pair consistent with originating from a
$Z$-boson (i.e. satisfying Eq.~(\ref{eq:mjj})) and which is not tagged
as a $b$-pair. This explicitly removes $Z\to b\bar b$ decays, 
reducing the combinatorial background and simplifying the analysis. Since
additional jets tend to weaken the sensitivity limits for
$\lambda$~\cite{tim}, I require exactly two light jets, and four tagged
$b$-quarks in events. Two or more pairs of $b$-quarks have to satisfy
Eq.~(\ref{eq:mbb}) or~(\ref{eq:mbb1}).

In addition to the four Feynman diagrams contributing to the ${\cal
O}(\alpha^3)$ $e^+e^-\to ZHH\to jjb\bar bb\bar b$ signal, there are
approximately 8,500 ${\cal O}(\alpha^6)$, ${\cal O}(\alpha_s^4\alpha^2)$ 
and ${\cal O}(\alpha_s^2\alpha^4)$ non-resonant and single Higgs
resonant diagrams contributing to the same final state. Single-resonant
Higgs and the ${\cal O}(\alpha_s^4\alpha^2)$ diagrams constitute the
potentially largest background contributions. In Ref.~\cite{LC_HH2}, the
role of the non-resonant and single resonant ${\cal
O}(\alpha_s^2\alpha^2)$ and ${\cal 
O}(\alpha^4)$ diagrams in $e^+e^-\to ZHH\to Zb\bar bb\bar b$ was analyzed
in the limit of a stable $Z$-boson. It was found that the contribution
of the ${\cal O}(\alpha_s^2\alpha^2)$ diagrams was small compared with
the signal 
for $m_H\leq 140$~GeV. However, the background from ${\cal O}(\alpha^4)$
single Higgs production can be substantial, in particular for
$m_H>130$~GeV. 
Misidentification of light jets and charm quarks also contribute to the
background for $jjb\bar bb\bar b$ production. 

Using the cuts and efficiencies listed in Eqs.~(\ref{eq:cuts1})
and~(\ref{eq:eff1}), and requiring one light jet pair satisfying 
Eq.~(\ref{eq:mjj}) and four tagged $b$-jets with at least two pairs
fulfilling Eq.~(\ref{eq:mbb}), I calculate the cross
section of the $ZHH\to jjb\bar bb\bar b$ signal, the cross section for
$e^+e^-\to jjb\bar bb\bar b$ including all ${\cal O}(\alpha^6)$, ${\cal
O}(\alpha_s^4\alpha^2)$  
and ${\cal O}(\alpha_s^2\alpha^4)$ diagrams, the $e^+e^-\to jjb\bar
bc\bar c$ background (approximately 7,300 ${\cal O}(\alpha^6)$, ${\cal
O}(\alpha_s^4\alpha^2)$ and ${\cal O}(\alpha_s^2\alpha^4)$ diagrams),
the $b\bar b 4j$ background (approximately 15,600 ${\cal O}(\alpha^6)$, ${\cal
O}(\alpha_s^4\alpha^2)$ and ${\cal O}(\alpha_s^2\alpha^4)$ diagrams),
and the ${\cal O}(\alpha_s^4\alpha^2)$ $e^+e^-\to 6j$ background
(approximately 7,600 diagrams). The signal is calculated for the SM
Higgs self-coupling ($\Delta\lambda_{HHH}=0$), $\Delta\lambda_{HHH}=+1$,
and $\Delta\lambda_{HHH}=-1$, where
\begin{equation}
\label{eq:lam}
\Delta\lambda_{HHH}=\lambda_{HHH}-1=\frac{\lambda}{\lambda_{SM}}-1.
\end{equation}

\begin{figure}[th!] 
\begin{center}
\includegraphics[width=10.9cm]{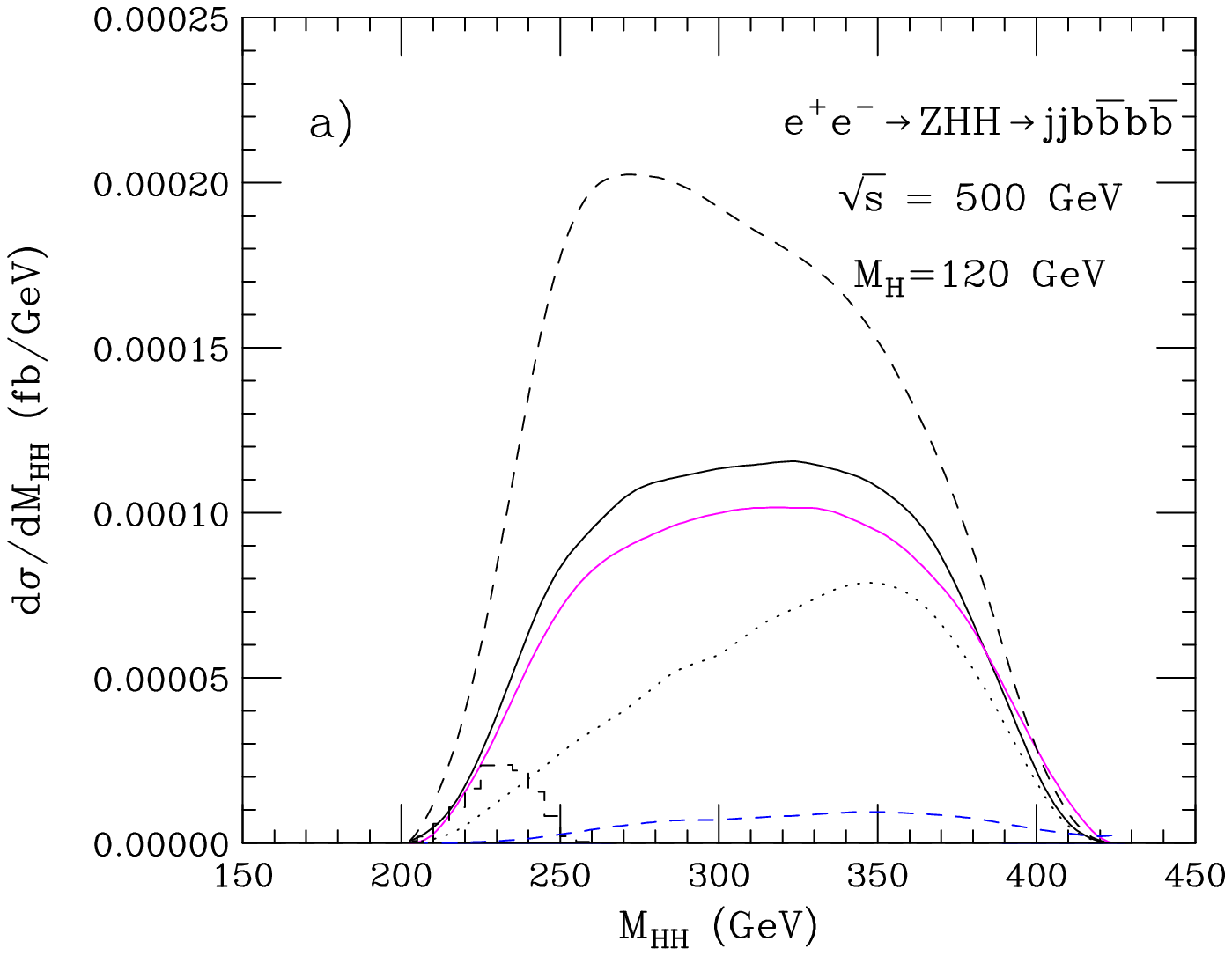} \\[3mm]
\includegraphics[width=10.9cm]{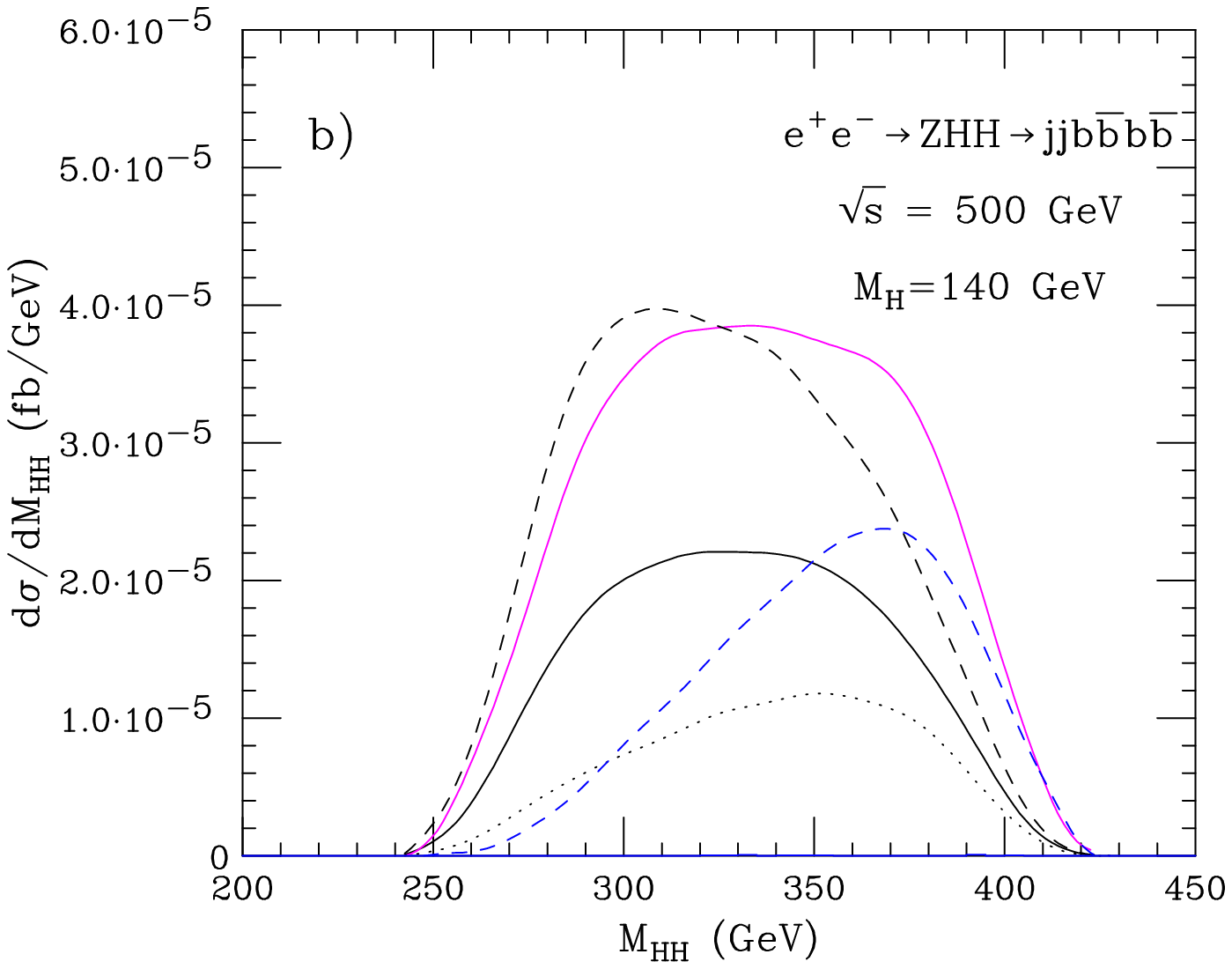}
\vspace*{2mm}
\caption[]{\label{fig:one} 
The $e^+e^-\to ZHH\to jjb\bar bb\bar b$ cross section with four $b$-tags
as a function of
the Higgs pair invariant mass, $M_{HH}$, for $\sqrt{s}=0.5$~TeV and a)
$m_H=120$~GeV and b) $m_H=140$~GeV. The black solid line is the
prediction of the SM signal cross section. The black dashed and dotted
lines correspond to the $ZHH$ signal cross section for
$\Delta\lambda_{HHH}=+1$ and $\Delta\lambda_{HHH}=-1$, respectively. The
dashed histogram in part a) represents the combinatorial background from
incorrectly assigned $b\bar b$ pairs in the signal. The magenta line
shows the SM cross section for $e^+e^-\to jjb\bar bb\bar b$ including the
full set of ${\cal O}(\alpha^6)$, ${\cal O}(\alpha_s^4\alpha^2)$ 
and ${\cal O}(\alpha_s^2\alpha^4)$ Feynman diagrams. The dashed blue
line corresponds to the $jjb\bar bc\bar c$ cross section. The cuts
imposed and the efficiencies used are summarized in Eqs.~(\ref{eq:cuts1}),
(\ref{eq:eff1}), and (\ref{eq:mjj}) --~(\ref{eq:mbb1}). }
\vspace{-7mm}
\end{center}
\end{figure}

\begin{figure}[th!] 
\begin{center}
\includegraphics[width=11.3cm]{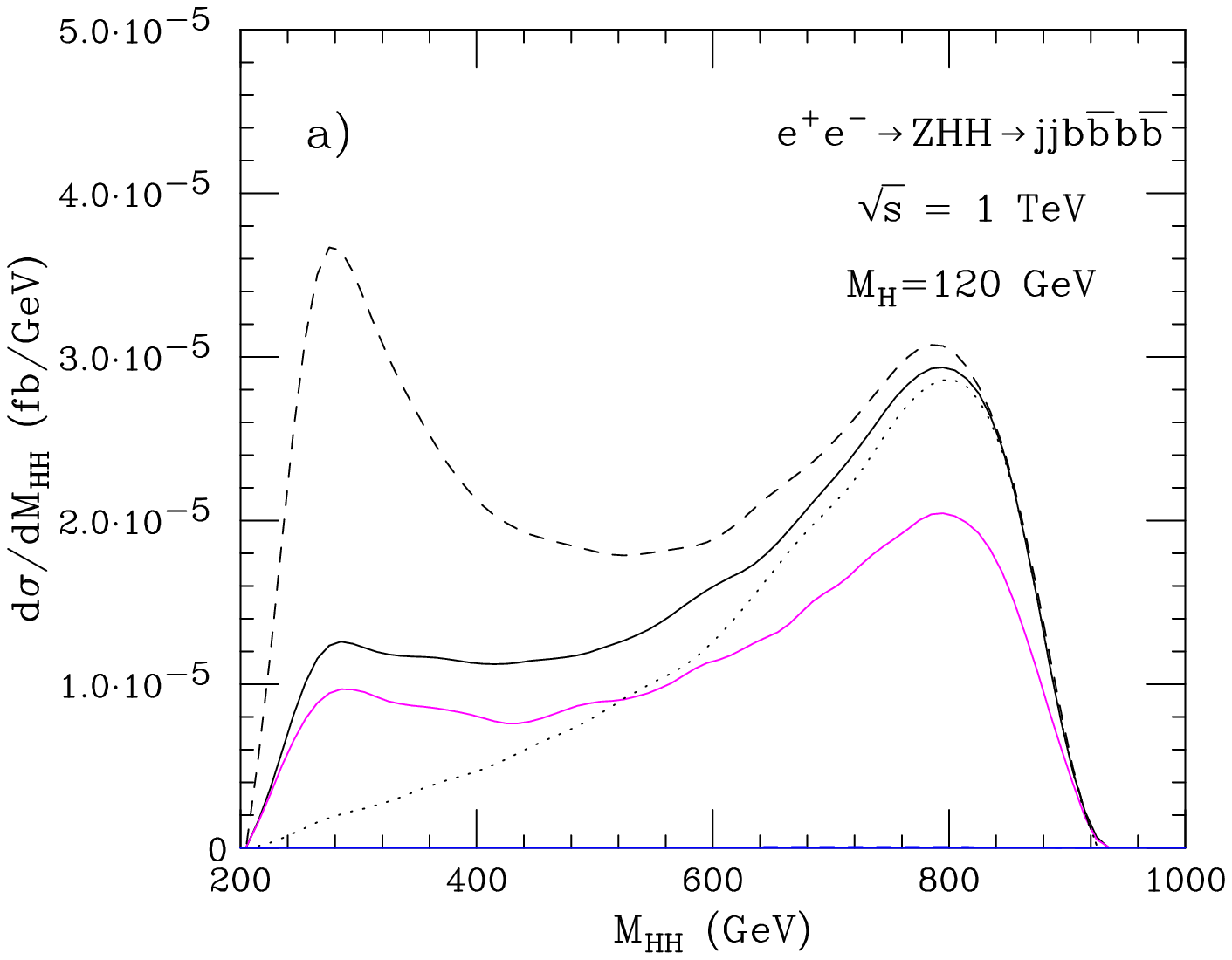} \\[3mm]
\includegraphics[width=11.3cm]{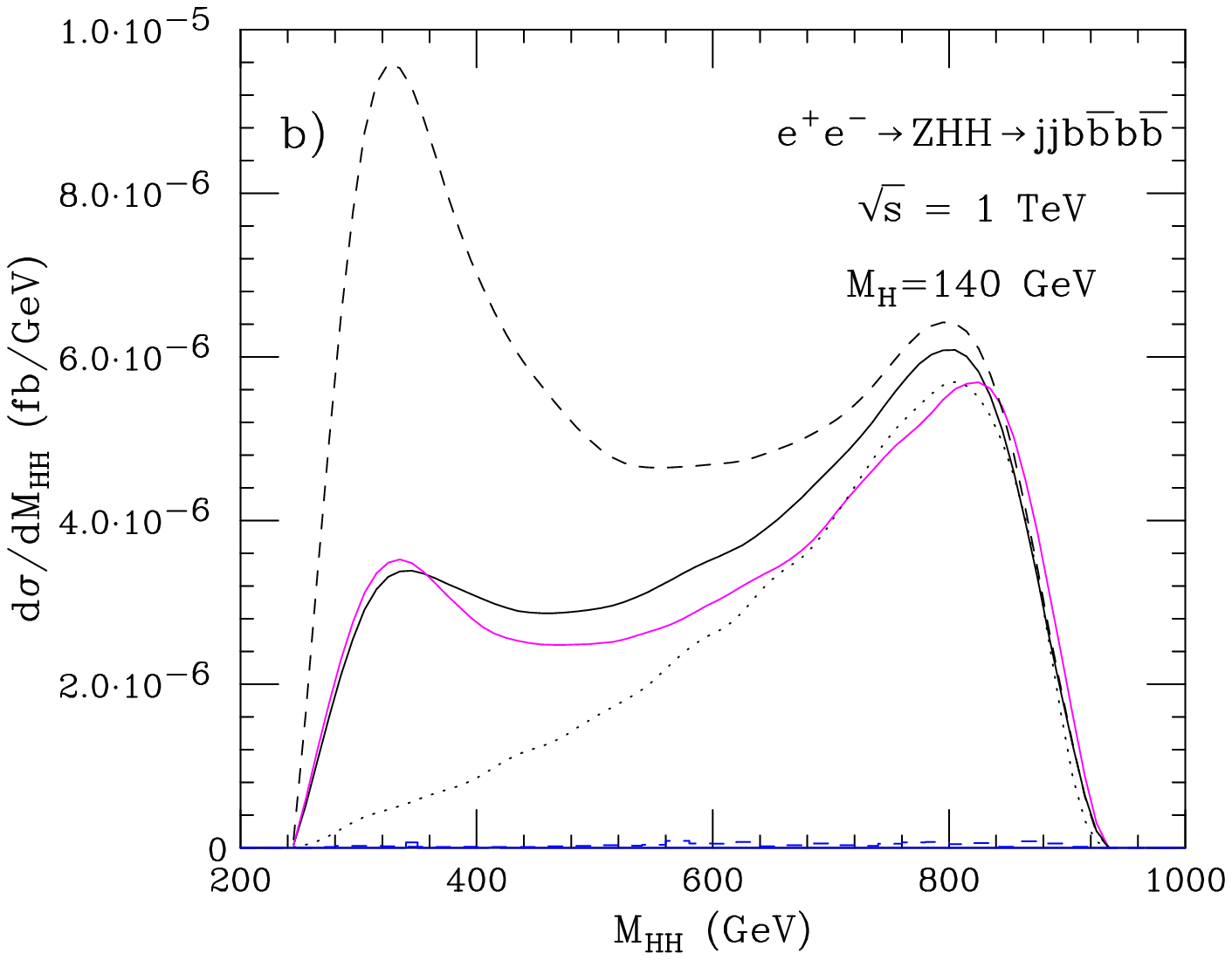}
\vspace*{2mm}
\caption[]{\label{fig:two} 
The $e^+e^-\to ZHH\to jjb\bar bb\bar b$ cross section with four $b$-tags
as a function of
the Higgs pair invariant mass, $M_{HH}$, for $\sqrt{s}=1$~TeV and a)
$m_H=120$~GeV and b) $m_H=140$~GeV. The black solid line is the
prediction of the SM signal cross section. The black dashed and dotted
lines correspond to the $ZHH$ signal cross section for
$\Delta\lambda_{HHH}=+1$ and $\Delta\lambda_{HHH}=-1$, respectively. The
magenta line 
shows the SM cross section for $e^+e^-\to jjb\bar bb\bar b$ including the
full set of ${\cal O}(\alpha^6)$, ${\cal O}(\alpha_s^4\alpha^2)$ 
and ${\cal O}(\alpha_s^2\alpha^4)$ Feynman diagrams. The dashed blue
line corresponds to the $jjb\bar bc\bar c$ cross section. The cuts
imposed and the efficiencies used are summarized in Eqs.~(\ref{eq:cuts1}),
(\ref{eq:eff1}), and (\ref{eq:mjj}) --~(\ref{eq:mbb1}). }
\vspace{-7mm}
\end{center}
\end{figure}

The results for the Higgs pair invariant mass distribution with
$\sqrt{s}=0.5$~TeV (1~TeV), and $m_H=120$~GeV and 
$m_H=140$~GeV are shown in Fig.~\ref{fig:one} (Fig.~\ref{fig:two}). 
The solid black lines show the prediction for the SM $e^+e^-\to ZHH\to
jjb\bar bb\bar b$ signal. Since it is impossible to know which $b$-quark
has to be 
paired with which $\bar b$-quark when reconstructing the Higgs bosons in
the event, there is a combinatorial background from incorrect pairing,
which is largest close to threshold. The dashed histogram in
Fig.~\ref{fig:one}a shows that the combinatorial background indeed
dominates the signal cross section close to threshold, but dies out
quickly for larger values of $M_{HH}$. Similar results are obtained for
different Higgs boson masses and center of mass energies. 

The dashed and dotted lines give the $M_{HH}$ distribution for
$\Delta\lambda_{HHH}=+1$ and $\Delta\lambda_{HHH}=-1$,
respectively. They demonstrate that the Higgs pair invariant mass
distribution is sensitive to the Higgs self-coupling $\lambda$,
especially for small values of $M_{HH}$. Positive (negative) values of
$\Delta\lambda_{HHH}$ increase (decrease) the $ZHH$ cross section. The
magenta curves give the SM cross section for $e^+e^-\to jjb\bar bb\bar
b$ including the 
full set of ${\cal O}(\alpha^6)$, ${\cal O}(\alpha_s^4\alpha^2)$ 
and ${\cal O}(\alpha_s^2\alpha^4)$ Feynman diagrams. For $m_H=120$~GeV
and $\sqrt{s}=0.5$~TeV and the renormalization scale set to $\mu=M_Z$,
the contribution from the single-resonant and non-resonant electroweak 
and QCD diagrams, and the interference effects between these diagrams and
the $ZHH$ signal diagrams, 
decrease the cross section for $jjb\bar bb\bar b$ production by
about 10\%. For $m_H=140$~GeV at the same center of mass energy,
however, taking into account the single-resonant and non-resonant
electroweak and QCD diagrams almost doubles the cross section. Since
these diagrams do not depend on the Higgs self-coupling, this
considerably reduces the sensitivity to $\lambda$ for
$m_H=140$~GeV. Due to the reduced phase space and the smaller $H\to b\bar b$
branching ratio, the $ZHH\to jjb\bar bb\bar b$ cross section for
$m_H=140$~GeV is 
about a factor~8 smaller than that for $m_H=120$~GeV. As I will show
in Sec.~\ref{sec:six}, the reduction in signal cross section, combined
with the increase of the single-resonant and non-resonant background,
will make it very difficult to measure the Higgs self-coupling in
$jjb\bar bb\bar b$ production for
$m_H>120$~GeV at an $e^+e^-$ collider with $\sqrt{s}=500$~GeV. 

The contribution of the non-resonant diagrams to the $jjb\bar bb\bar b$
cross section can, in principle, be reduced by imposing a tighter cut on
the $b\bar b$ invariant mass. However, the energy loss of $b$-quarks,
combined with the finite resolution of detectors smears out the Higgs boson
resonance over a fairly large $b\bar b$ invariant mass range. A more
stringent $m(b\bar b)$ cut thus would, at the same time, considerably
reduce the signal cross section.

For $\sqrt{s}=1$~TeV and $m_H=120$~GeV ($m_H=140$~GeV), the
single-resonant and non-resonant background diagrams lower the cross
section by up to 50\% (20\%). The cross section reduction is mainly
caused by the interference of resonant $Z\to jj$, and non-resonant
diagrams. Note that the effect of the single- and non-resonant
background can easily be confused with that of a non-SM Higgs
self-coupling, especially for a small number of signal events. At
$\sqrt{s}=1$~TeV, the reducible backgrounds are almost negligible. The
SM cross section peaks in the region which is least sensitive to
$\lambda_{HHH}$. Given a sufficient integrated luminosity, this makes it
possible to use the high $M_{HH}$ region to normalize the cross
section. Unfortunately, this method does not work for
$\sqrt{s}=0.5$~TeV, where non-standard Higgs self-couplings lead to a
broad enhancement or reduction of the cross section, spread out over
most of the accessible $M_{HH}$ range.

The largest reducible background to $ZHH\to jjb\bar bb\bar b$ originates from
$jjb\bar bc\bar c$ production where both $c$-quarks are mistagged as
$b$-jets. This background is only significant for $\sqrt{s}=0.5$~TeV and
$m_H=140$~GeV. The $b\bar b4j$ background where two light jets are
misidentified as $b$-jets, and the $e^+e^-\to 6j$ background where four
jets are mistagged, are very small and are not shown in
Figs.~\ref{fig:one} and~\ref{fig:two}.

\begin{figure}[th!] 
\begin{center}
\includegraphics[width=11.2cm]{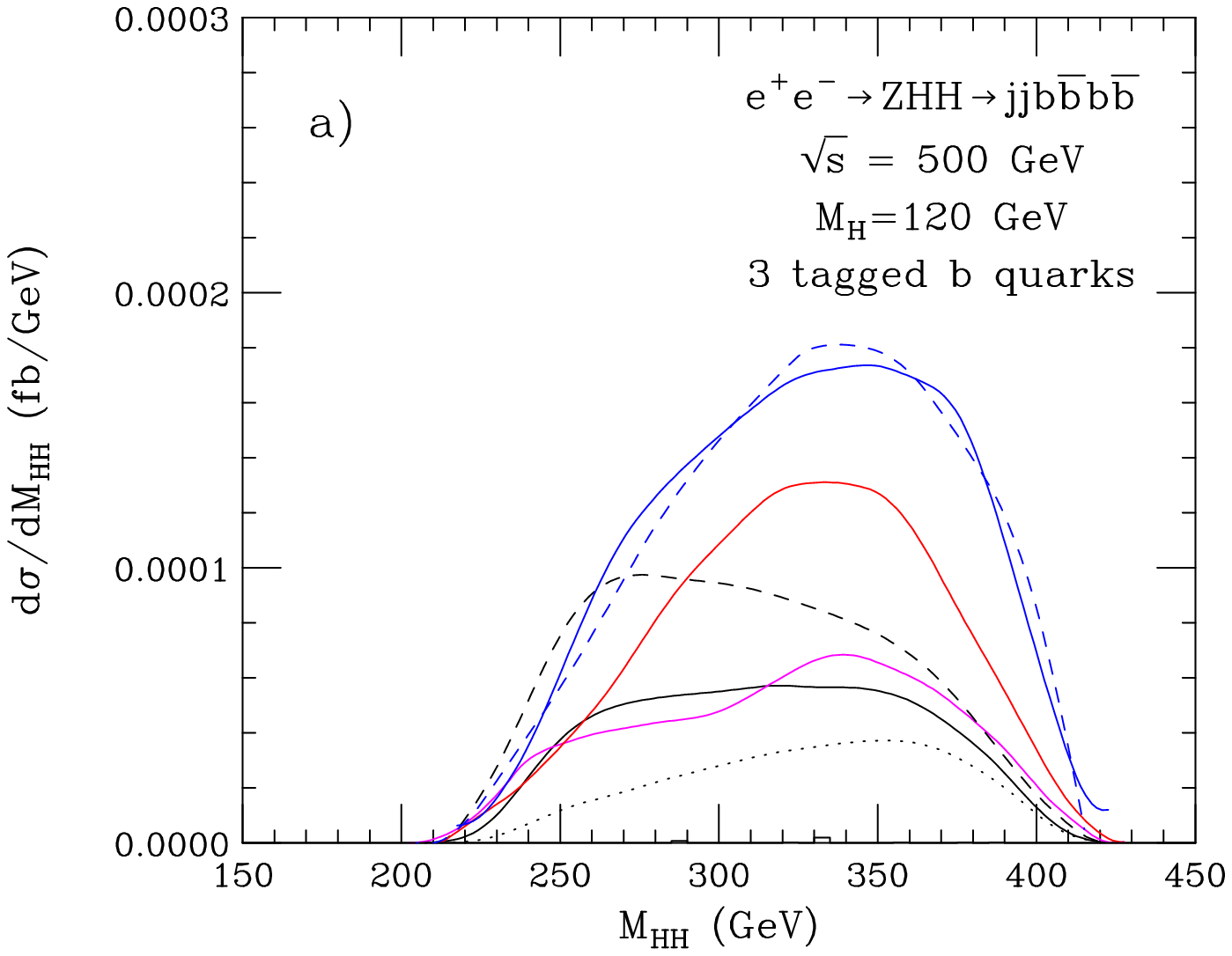} \\[3mm]
\includegraphics[width=11.2cm]{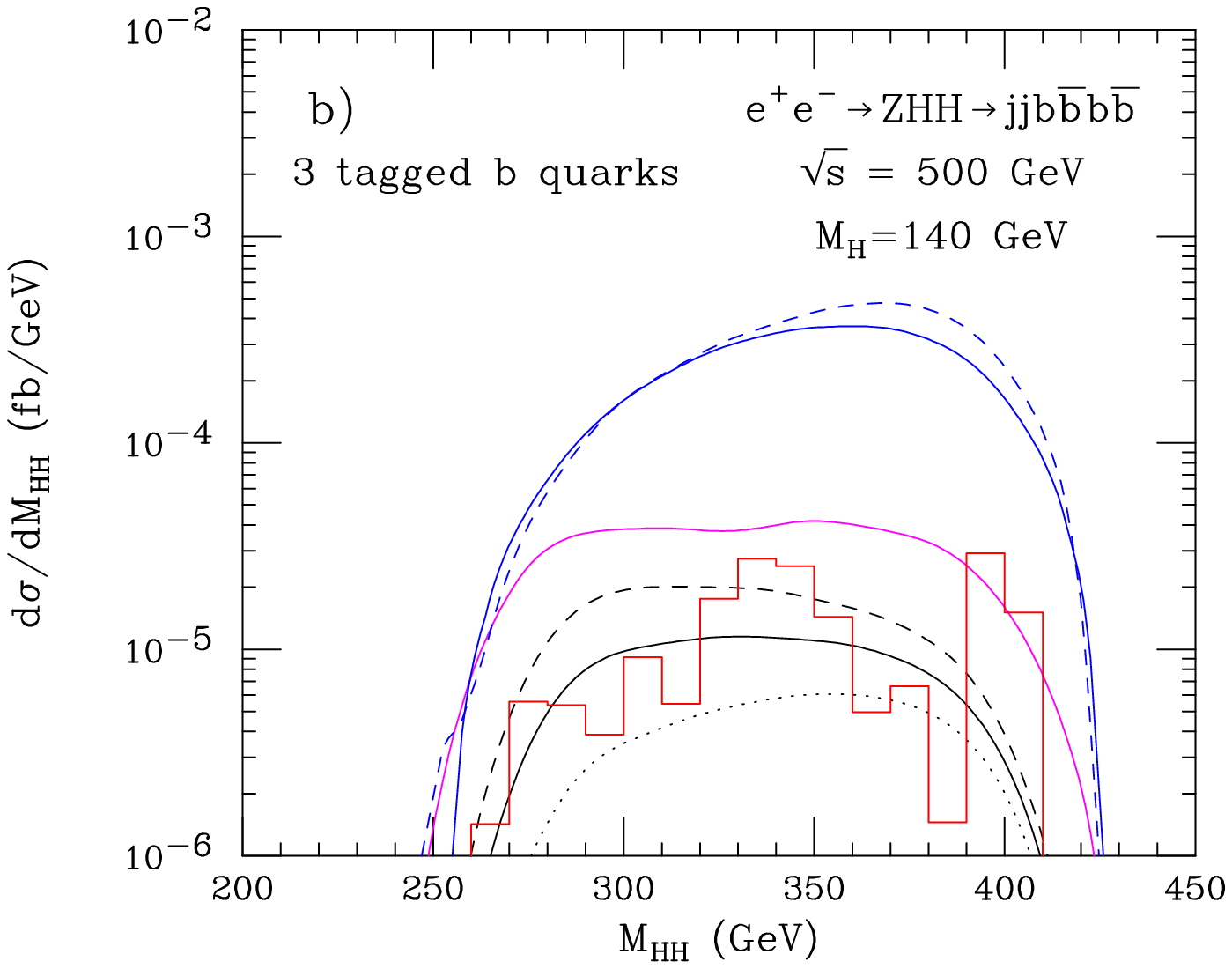}
\vspace*{2mm}
\caption[]{\label{fig:three} 
The $e^+e^-\to ZHH\to jjb\bar bb\bar b$ cross section with three
$b$-tags as a function of
the Higgs pair invariant mass, $M_{HH}$, for $\sqrt{s}=0.5$~TeV and a)
$m_H=120$~GeV and b) $m_H=140$~GeV. The black solid line is the
prediction of the SM signal cross section. The black dashed and dotted
lines correspond to the $ZHH$ signal cross section for
$\Delta\lambda_{HHH}=+1$ and $\Delta\lambda_{HHH}=-1$, respectively. The
magenta line 
shows the SM cross section for $e^+e^-\to jjb\bar bb\bar b$ including the
full set of ${\cal O}(\alpha^6)$, ${\cal O}(\alpha_s^4\alpha^2)$ 
and ${\cal O}(\alpha_s^2\alpha^4)$ Feynman diagrams. The solid (dashed) blue
and red lines or histograms correspond to the $b\bar bcjjj$ ($jjb\bar
bc\bar c$) and $b\bar b4j$ cross section, respectively. The cuts 
imposed and the efficiencies used are summarized in Eqs.~(\ref{eq:cuts1}),
(\ref{eq:eff1}), and (\ref{eq:mjj}) --~(\ref{eq:mbb1}). }
\vspace{-7mm}
\end{center}
\end{figure}

\begin{figure}[th!] 
\begin{center}
\includegraphics[width=11.2cm]{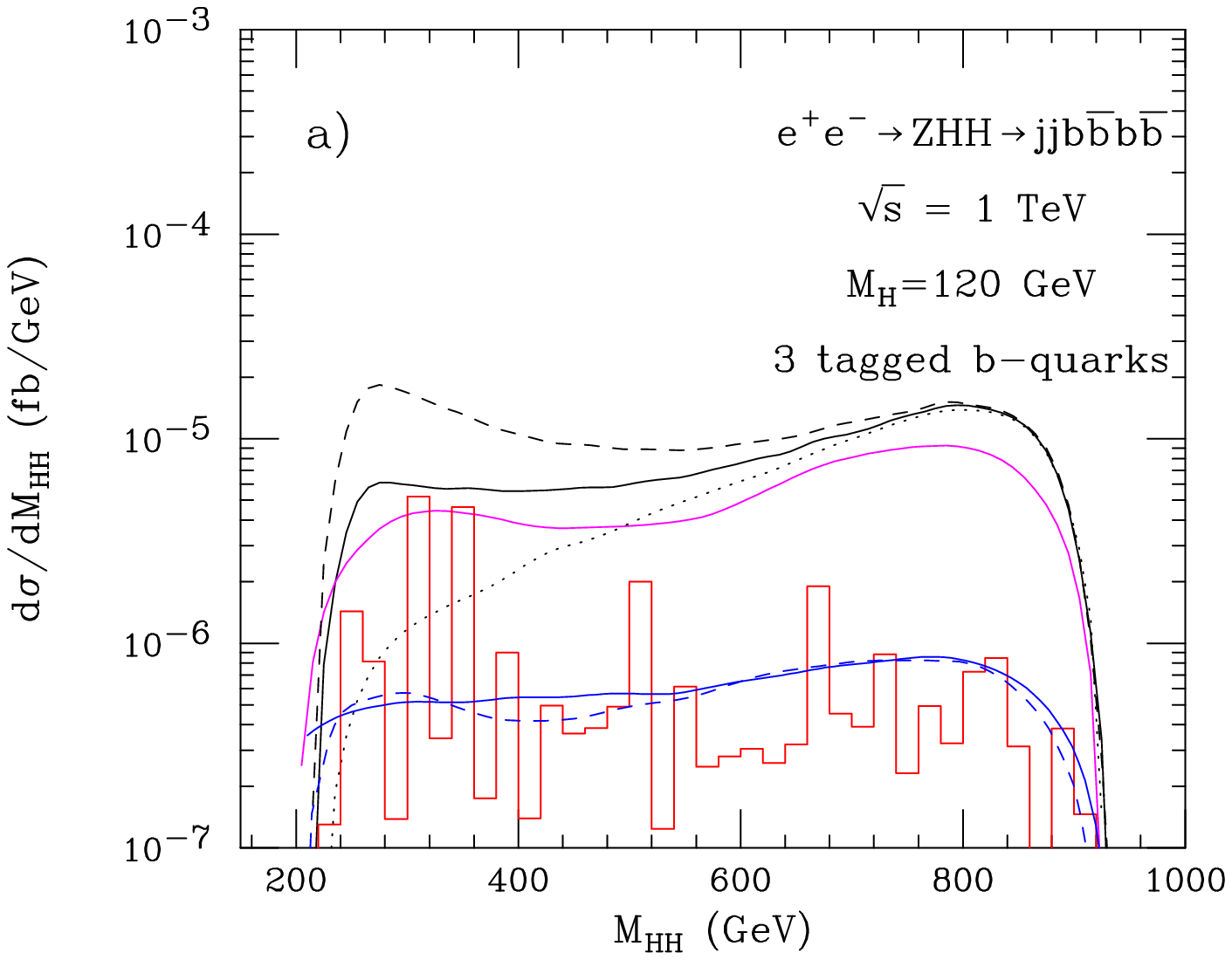} \\[3mm]
\includegraphics[width=11.2cm]{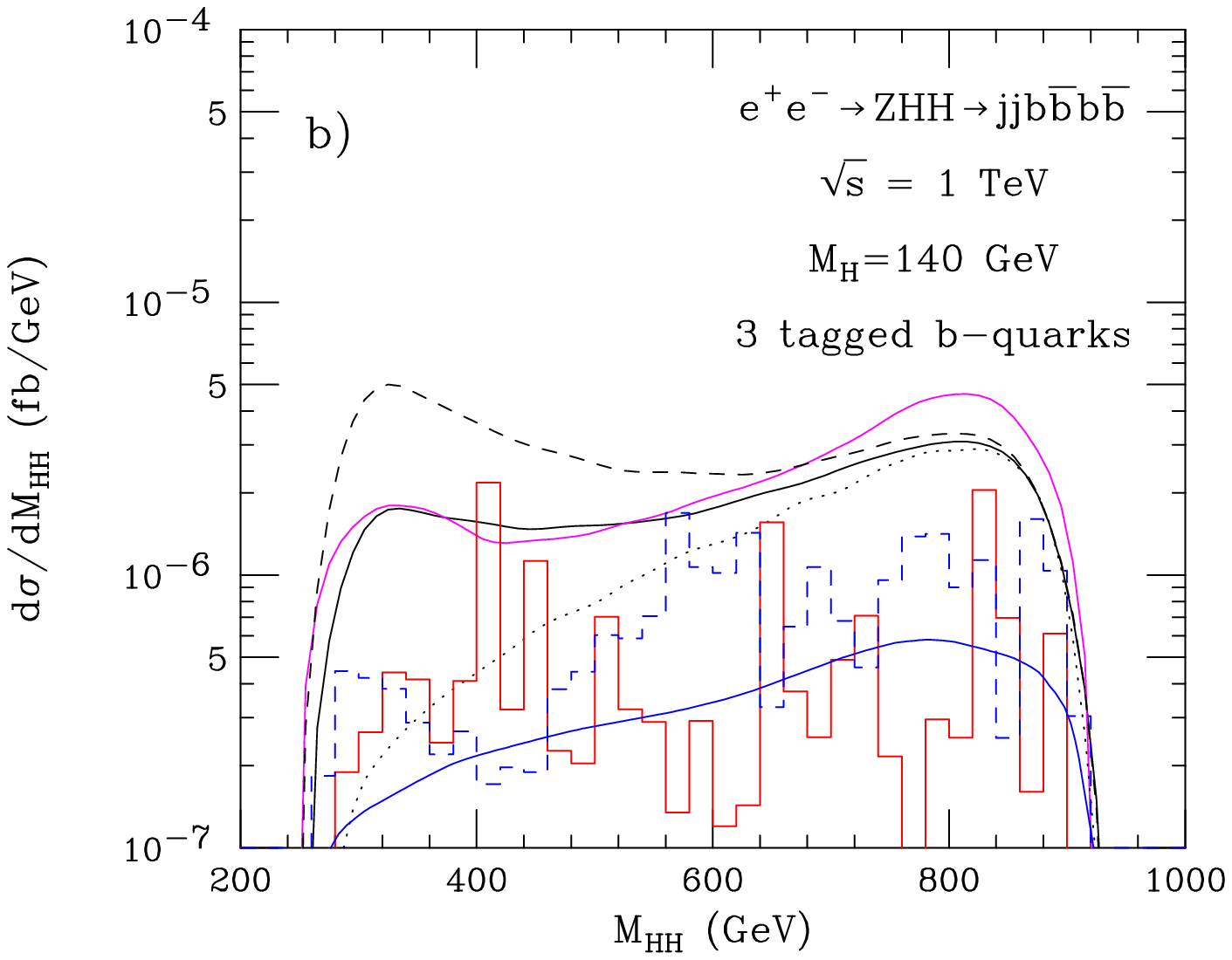}
\vspace*{2mm}
\caption[]{\label{fig:four} 
The $e^+e^-\to ZHH\to jjb\bar bb\bar b$ cross section with three
$b$-tags as a function of
the Higgs pair invariant mass, $M_{HH}$, for $\sqrt{s}=1$~TeV and a)
$m_H=120$~GeV and b) $m_H=140$~GeV. The black solid line is the
prediction of the SM signal cross section. The black dashed and dotted
lines correspond to the $ZHH$ signal cross section for
$\Delta\lambda_{HHH}=+1$ and $\Delta\lambda_{HHH}=-1$, respectively. The
magenta line 
shows the SM cross section for $e^+e^-\to jjb\bar bb\bar b$ including the
full set of ${\cal O}(\alpha^6)$, ${\cal O}(\alpha_s^4\alpha^2)$ 
and ${\cal O}(\alpha_s^2\alpha^4)$ Feynman diagrams. The solid (dashed) blue
and red lines or histograms correspond to the $b\bar bcjjj$ ($jjb\bar
bc\bar c$) and $b\bar b4j$ cross section, respectively. The cuts 
imposed and the efficiencies used are summarized in Eqs.~(\ref{eq:cuts1}),
(\ref{eq:eff1}), and (\ref{eq:mjj}) --~(\ref{eq:mbb1}). }
\vspace{-7mm}
\end{center}
\end{figure}

Although the expected $b$-tagging efficiency at the ILC is very high,
requiring four tagged $b$-quarks reduces the observable cross section by
a factor 0.66 for $\epsilon_b=0.9$. Since the signal cross section is
very small, it is natural to explore whether it is advantageous to reduce the
number of required $b$-tagged jets. 
Reducing the number of required $b$-tags from four to three increases
the signal cross section by a factor
\begin{equation}
1+\frac{4(1-\epsilon_b)}{\epsilon_b}
\end{equation}
ie. for $\epsilon_b=0.9$ by about a factor 1.44. However, the $jjb\bar
bc\bar c$ background increases by about a factor~10, while the $b\bar
b4j$ and $6j$ backgrounds grow by more than two orders of
magnitude. Furthermore, $b\bar bcjjj$ production now also contributes to
the background. The $M_{HH}$ distribution for $ZHH\to jjb\bar bb\bar b$
with three tagged $b$-quarks and $\sqrt{s}=0.5$~TeV (1~TeV) is shown in
Fig.~\ref{fig:three} (Fig.~\ref{fig:four}). 
To calculate the cross sections shown in these figures, I require one
light jet pair in the mass window given in Eq.~(\ref{eq:mjj}), one
tagged $b\bar b$ pair and at least one $bj$ combination with an
invariant mass satisfying Eq.~(\ref{eq:mbb}) or~(\ref{eq:mbb1}).
For a center of mass energy of 0.5~TeV, the $jjb\bar bc\bar c$ (dashed
blue line or histogram), $b\bar bcjjj$ (solid blue line or histogram),
and the $b\bar b4j$ background (solid red line or histogram) are all
significantly larger than the $ZHH$ signal cross section. At an $e^+e^-$
collider with $\sqrt{s}=1$~TeV, on the other hand, these backgrounds are
at most of the size of the signal. Thus, requiring three or more
$b$-tags may improve the sensitivity to $\lambda$ at a 1~TeV $e^+e^-$
collider. For $\sqrt{s}=0.5$~TeV, the dramatically
increased background may well lead to a decreased sensitivity, despite
the gain in the signal cross section.

The results shown in Figs.~\ref{fig:three} and~\ref{fig:four} assume the
efficiencies and misidentification probabilities of
Eq.~(\ref{eq:eff1}). A reduction of a factor~5 or more in most
backgrounds can be achieved if one is willing to accept a slightly
reduced $b$-tagging efficiency; see Eq.~(\ref{eq:eff2}). A
detailed analysis of the sensitivity limits which one may hope to
achieve requiring four tagged $b$-quarks, or three or more tagged $b$'s
using the efficiencies of Eq.~(\ref{eq:eff1}) or Eq.~(\ref{eq:eff2})
will be presented in Sec.~\ref{sec:six}. The cross sections for the two
sets of efficiencies and misidentification probabilities, and the SM
signal, $e^+e^-\to jjb\bar bb\bar b$ including the full set of ${\cal
O}(\alpha^6)$, ${\cal O}(\alpha_s^4\alpha^2)$ and ${\cal
O}(\alpha_s^2\alpha^4)$ Feynman diagrams, and the backgrounds are listed
in Table~\ref{tab:one} for three and four tagged $b$-jets, together with
the cross section for $\geq 3$ $b$-tags.
\begin{table}
\caption[]{\label{tab:one}
Cross sections in fb for $e^+e^-\to jjb\bar bb\bar b$ for $m_H=120$~GeV and
$m_H=140$~GeV, and $\sqrt{s}=0.5$~TeV and
$\sqrt{s}=1$~TeV. Shown are the results for the SM signal, the full set
of ${\cal O}(\alpha^6)$, ${\cal O}(\alpha_s^4\alpha^2)$ and ${\cal
O}(\alpha_s^2\alpha^4)$ Feynman diagrams (labeled ``all''), and the
total reducible backgrounds (labeled ``bgd''), with 3, 4 
and $\geq 3$ $b$-tags. Cross sections are listed for two sets of
$b$-tagging efficiencies and light quark/gluon jet and charm quark
misidentification probabilities (see Eqs.~(\ref{eq:eff1})
and~(\ref{eq:eff2})).} 
\vspace{2mm}
\begin{tabular}{c|ccc|ccc|ccc}
\multicolumn{10}{c}{$\epsilon_b=0.9$, $P_{c\to b}=0.1$, $P_{j\to
b}=0.005$}\\
\hline
\phantom{i} & \multicolumn{3}{c|}{4 $b$-tags} &  \multicolumn{3}{c|}{3
$b$-tags} &  \multicolumn{3}{c}{$\geq 3$ $b$-tags} \\
\phantom{j} & signal & all & bgd & signal & all & bgd & signal & all & bgd \\
\hline
$\sqrt{s}=0.5$~TeV, $m_H=120$~GeV & 0.016 & 0.014 & 0.001 & 0.008 &
0.008 & 0.061 & 0.024 & 0.022 & 0.062 \\ 
$\sqrt{s}=0.5$~TeV, $m_H=140$~GeV & 0.002 & 0.004 & 0.002 & 0.001 &
0.005 & 0.077 & 0.003 & 0.009 & 0.079 \\
\hline
$\sqrt{s}=1$~TeV, $m_H=120$~GeV & 0.011 & 0.008 & $2\cdot 10^{-5}$ & 0.005 &
0.004 & 0.001 & 0.016 & 0.013 & 0.001 \\ 
$\sqrt{s}=1$~TeV, $m_H=140$~GeV & 0.002 & 0.002 & $2\cdot 10^{-5}$ & 0.0012 &
0.0015 & 0.0010 & 0.003 & 0.004 & 0.001\\
\hline
\multicolumn{10}{c}{$\epsilon_b=0.8$, $P_{c\to b}=0.02$, $P_{j\to
b}=0.001$}\\
\hline
\phantom{i} & \multicolumn{3}{c|}{4 $b$-tags} &  \multicolumn{3}{c|}{3
$b$-tags} &  \multicolumn{3}{c}{$\geq 3$ $b$-tags} \\
\phantom{j} & signal & all & bgd & signal & all & bgd & signal & all & bgd \\
\hline
$\sqrt{s}=0.5$~TeV, $m_H=120$~GeV & 0.010 & 0.009 & $3\cdot 10^{-5}$ & 0.011 &
0.011 & 0.010 & 0.021 & 0.020 & 0.010 \\ 
$\sqrt{s}=0.5$~TeV, $m_H=140$~GeV & 0.001 & 0.002 & $6\cdot 10^{-5}$ & 0.001 &
0.007 & 0.012 & 0.002 & 0.009 & 0.012 \\
\hline
$\sqrt{s}=1$~TeV, $m_H=120$~GeV & 0.007 & 0.005 & 0 & 0.007 &
0.006 & $2\cdot 10^{-4}$ & 0.014 & 0.011 & $2\cdot 10^{-4}$ \\ 
$\sqrt{s}=1$~TeV, $m_H=140$~GeV & 0.001 & 0.001 & 0 & 0.0017 &
0.0021 & $2\cdot 10^{-4}$ & 0.003 & 0.003 & $2\cdot 10^{-4}$
\end{tabular}
\end{table}

One may wonder to what extent the renormalization scale uncertainty
affects the cross sections listed in Table~\ref{tab:one}. The reducible
backgrounds are all of ${\cal O}(\alpha_s^4\alpha^2)$ and thus are
uncertain by a factor $2^{\pm 1}$ or so. In the $e^+e^-\to jjb\bar
bb\bar b$ cross section which has been calculated using the full set of
${\cal O}(\alpha^6)$, ${\cal O}(\alpha_s^4\alpha^2)$ and ${\cal 
O}(\alpha_s^2\alpha^4)$ Feynman diagrams, the  ${\cal
O}(\alpha_s^4\alpha^2)$ diagrams play an important role only for
$\sqrt{s}=0.5$~TeV and $m_H=140$~GeV. In this case, I estimate that the
scale uncertainty may change the cross section by a factor $1.5^{\pm
1}$.

Table~\ref{tab:one} shows that, for $m_H=120$~GeV, $jjb\bar bb\bar b$
production 
with four tagged $b$-quarks provides a clean, albeit low statistics,
signal with a relatively 
small background. Including the final state with three $b$-tags
increases the signal cross section by almost 50\%. While the background
is still small for $\sqrt{s}=1$~TeV, it becomes substantial for
$\sqrt{s}=0.5$~TeV. The signal to background ratio for
$\sqrt{s}=0.5$~TeV may be improved by
choosing an optimal combination of $b$-tagging efficiency and light
quark/gluon jet and charm quark misidentification
probability~\cite{Boumediene:2008hs}. Furthermore, in absence of a
calculation of the NLO 
QCD corrections for $jjb \bar bc\bar c$, $b\bar bcjjj$ and $b\bar b4j$
production, the 
background cross section is subject to a large renormalization scale
uncertainty. For $m_H=140$~GeV, the signal cross section is so small
that even with an integrated luminosity of several ab$^{-1}$, only a
handful of signal events is produced. 

For $\sqrt{s}=1$~TeV, the
background is already small for the efficiencies listed in
Eq.~(\ref{eq:eff1}) and not much is gained by choosing a different
combination of $b$-tagging efficiency and light
quark/gluon jet and charm quark misidentification probability. 

The signal cross section may be further increased by relaxing the number
of $b$-tagged jets to two. In this case, one of the Higgs bosons may
undergo the decay $H\to c\bar c$ or $H\to gg$. For $m_H=120$~GeV, this
increases the signal 
cross section by about a factor~1.3. Unfortunately, the backgrounds increase
by a much larger factor, and overwhelm the signal~\cite{Timb}. I
therefore do not analyze the $ZHH\to b\bar b4j$ signal here. 

For $m_H\geq 140$~GeV, the relatively small branching ratio of $B(H\to
b\bar b)\approx 30\%$ makes it difficult to measure $\lambda$ in $jjb\bar
bb\bar b$
production, regardless of the $b$-tagging efficiency. In this Higgs mass
range, $B(H\to W^*W)\approx 50\%$ and final states such as $b\bar b6j$ and
$\ell^\pm\nu b\bar b4j$ offer the possibility to more than
double the number of observed $ZHH$ events. I do not investigate
these final states here. For $m_H>2M_W$, most Higgs bosons decay into a
pair of $W$ bosons. While there is not enough phase space for $ZHH$
production in this region at a 500~GeV $e^+e^-$ collider, a small number
of $ZHH$ events may be produced for $\sqrt{s}=1$~TeV or above. However,
as I will show in the following Section, the cross section for
$\nu\bar\nu HH$ production is considerably larger in this
region. $ZHH$ production, therefore, is not discussed here for $m_H>2M_W$.

\section{$\nu\bar\nu HH$ Analysis}
\label{sec:five}

For $\sqrt{s}\geq 1$~TeV, $e^+e^-\to\nu\bar\nu HH$
($\nu=\nu_e,\,\nu_\mu,\,\nu_\tau$) becomes the dominant 
source of Higgs boson pairs. For $\nu=\nu_\mu,\,\nu_\tau$ only $ZHH$
production with $Z\to\bar\nu_\mu\nu_\mu,\,\bar\nu_\tau\nu_\tau$
contributes. For $\nu=\nu_e$, a total of eight Feynman diagrams
contribute, including four $ZHH$, $Z\to\bar\nu_e\nu_e$ and four vector
boson fusion diagrams. In the following, I will discuss $\nu\bar\nu HH$
production and the relevant backgrounds for $m_H=120$~GeV, 140~GeV and
$m_H=180$~GeV, and $\sqrt{s}=1$~TeV and 3~TeV.

\subsection{$m_H=120$~GeV}
\label{sec:fivea}

For $m_H=120$~GeV, $H\to b\bar b$ decays dominate. I will therefore
concentrate on the $\nu\bar\nu b\bar bb\bar b$ final
state. Specifically, I require 
events with missing transverse momentum and four jets satisfying
Eq.~(\ref{eq:cuts1}). A minimum of three of the jets have to be tagged
as $b$-jets, and there have to be at least two jet pairs fulfilling
Eq.~(\ref{eq:mbb}). 

\begin{figure}[t!] 
\begin{center}
\includegraphics[width=15.5cm]{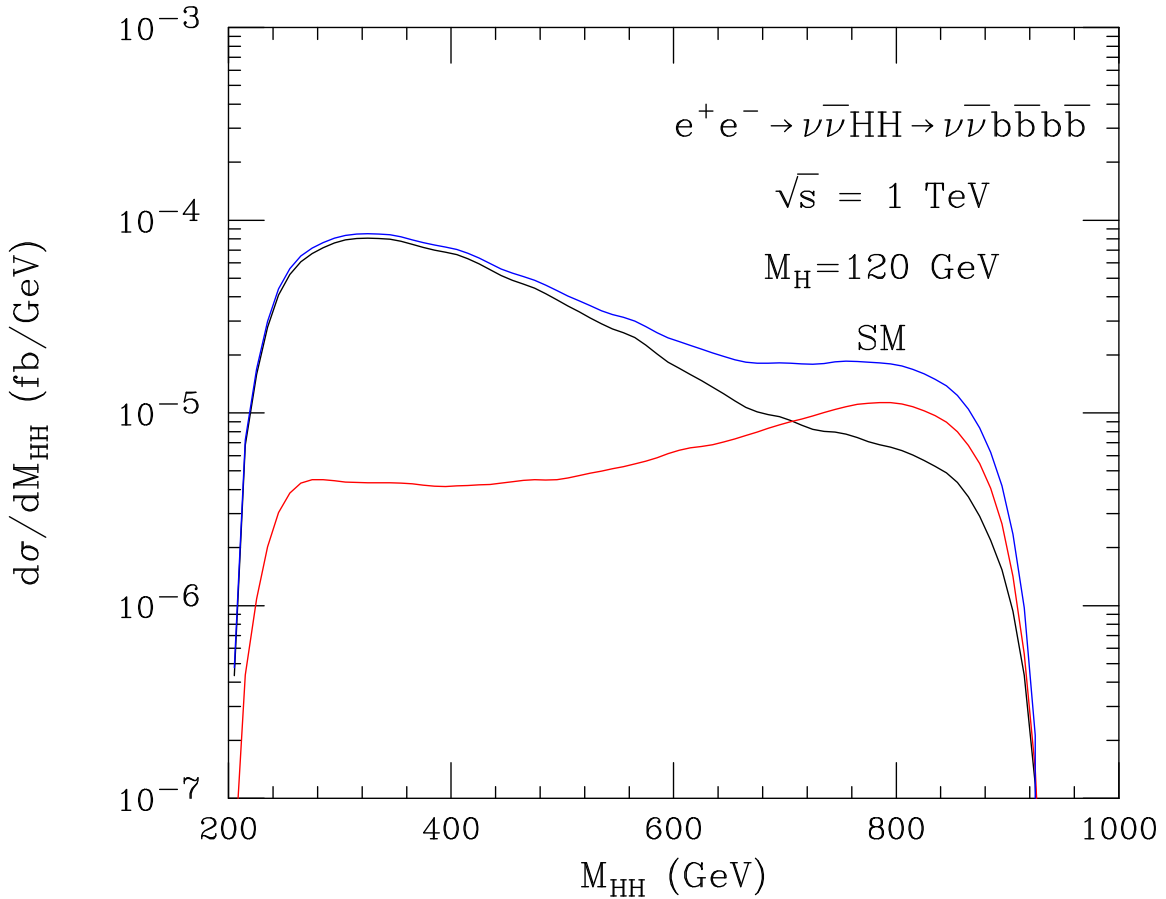}
\vspace*{2mm}
\caption[]{\label{fig:five} 
The SM $e^+e^-\to\nu\bar\nu HH\to\nu\bar\nu 4b$ cross section with three
$b$-tags as a function of
the Higgs pair invariant mass, $M_{HH}$, for $\sqrt{s}=1$~TeV. The black
(red) solid line shows the $ZHH\to\nu_l\bar\nu_l b\bar bb\bar b$
($l=\mu,\,\tau$) ($e^+e^-\to\nu_e\bar\nu_e HH\to\nu_e\bar\nu_e b\bar
bb\bar b$) cross section. The
blue curve represents the total $e^+e^-\to\nu\bar\nu HH\to\nu\bar\nu
b\bar bb\bar b$ cross section. The cuts 
imposed and the efficiencies used are summarized in Eqs.~(\ref{eq:cuts1}),
(\ref{eq:eff1}), (\ref{eq:mjj}), and~(\ref{eq:mbb}). }
\vspace{-7mm}
\end{center}
\end{figure}

The SM $ZHH\to\nu_l\bar\nu_l b\bar bb\bar b$ ($l=\mu,\,\tau$) and
$e^+e^-\to\nu_e\bar\nu_e HH\to\nu_e\bar\nu_e b\bar bb\bar b$ $M_{HH}$
differential cross sections for $\sqrt{s}=1$~TeV are shown by 
the solid black and red lines in Fig.~\ref{fig:five}. The blue line
gives the inclusive $e^+e^-\to\nu\bar\nu HH\to\nu\bar\nu b\bar bb\bar b$
cross section. 
For small values of $M_{HH}$, the cross section is completely dominated
by the process $e^+e^-\to\nu_e\bar\nu_e HH$, whereas for large Higgs
pair invariant masses, $ZHH$ production followed by $Z\to\nu_l\bar\nu_l$
gives the largest contribution to the inclusive $\nu\bar\nu HH$ cross
section. A qualitatively similar result is obtained at $\sqrt{s}=3$~TeV,
however, with the $e^+e^-\to\nu_e\bar\nu_e HH$ cross section being more
than a factor~100 larger than the $ZHH$ rate at small values of
$M_{HH}$. 

The main backgrounds to $\nu\bar\nu HH\to\nu\bar\nu b\bar bb\bar b$
production are $e^+e^-\to \nu\bar\nu b\bar bb\bar b$ at ${\cal
O}(\alpha_s^2\alpha^4)$ and ${\cal 
O}(\alpha^6)$ (about 2,300 single- and non-resonant Feynman diagrams),
and $\nu\bar\nu b\bar 
bc\bar c$ (about 900 ${\cal O}(\alpha_s^2\alpha^4)$ and ${\cal
O}(\alpha^6)$ Feynman diagrams) and $\nu\bar\nu b\bar bjj$ production
(about 2,100 ${\cal O}(\alpha_s^2\alpha^4)$ and ${\cal O}(\alpha^6)$
Feynman diagrams) where one or two charm or light quark/gluon jets are
mistagged. Potentially dangerous backgrounds also come from $b\bar
bb\bar b$ and 
$b\bar bjj$ production with the missing transverse originating from the
energy loss in $b$-decays or jet energy mismeasurements. 

\begin{figure}[th!] 
\begin{center}
\includegraphics[width=11.1cm]{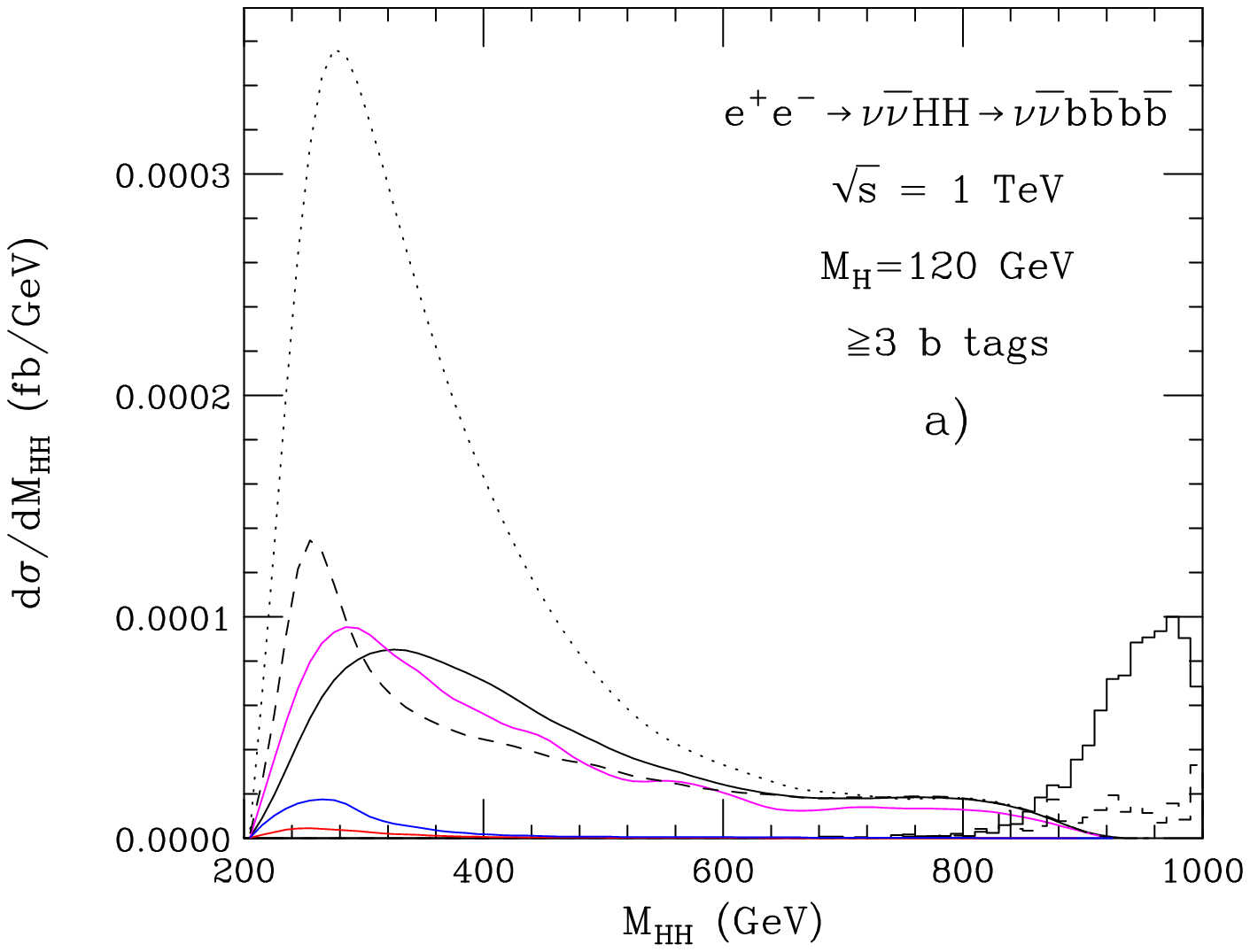} \\[3mm]
\includegraphics[width=11.1cm]{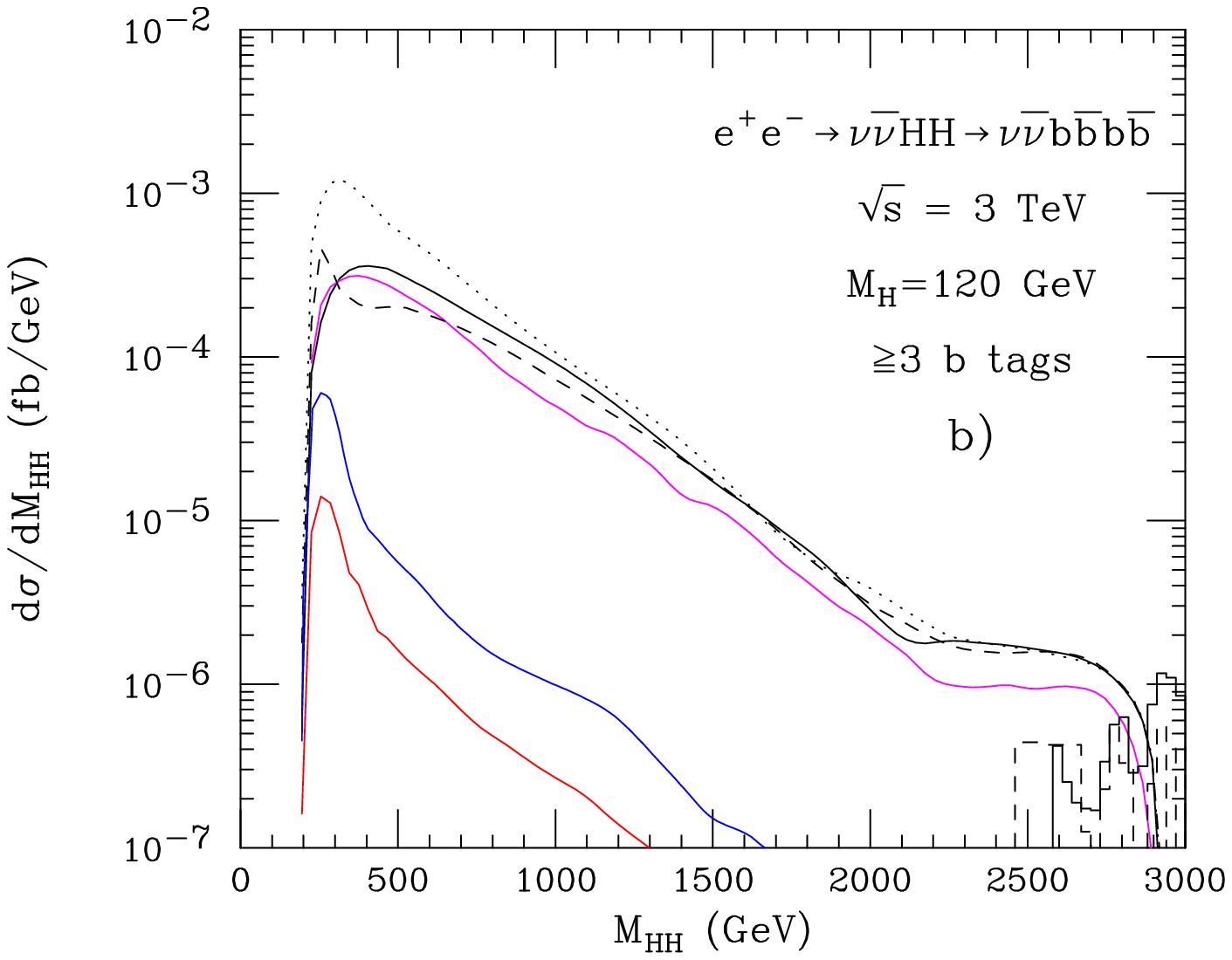}
\vspace*{2mm}
\caption[]{\label{fig:six} 
The $e^+e^-\to\nu\bar\nu HH\to\nu\bar\nu b\bar bb\bar b$ cross section
with three or more $b$-tags as a function of
the Higgs pair invariant mass, $M_{HH}$, for $m_H=120$~GeV and a)
$\sqrt{s}=1$~TeV and b) $\sqrt{s}=3$~TeV. 
The black solid line is the
prediction of the SM signal cross section. The black dashed and dotted
lines correspond to the $\nu\bar\nu HH$ signal cross section for
$\Delta\lambda_{HHH}=+1$ and $\Delta\lambda_{HHH}=-1$, respectively. The
magenta line 
shows the SM cross section for $e^+e^-\to\nu\bar\nu b\bar bb\bar b$
including the full set of ${\cal O}(\alpha^6)$ 
and ${\cal O}(\alpha_s^2\alpha^4)$ Feynman diagrams. The blue
and red lines correspond to the $\nu\bar\nu b\bar bc\bar c$ and
$\nu\bar\nu b\bar bjj$ cross section. The solid and dashed histograms
represent the prediction of the $b\bar bb\bar b$ and $b\bar bjj$
backgrounds. The cuts  
imposed and the efficiencies used are summarized in Eqs.~(\ref{eq:cuts1}),
(\ref{eq:eff1}), (\ref{eq:mjj}), and~(\ref{eq:mbb}). }
\vspace{-7mm}
\end{center}
\end{figure}

The Higgs-pair invariant mass distribution for $\sqrt{s}=1$~TeV and
3~TeV for the
SM $\nu\bar\nu HH$ signal (solid black curve), $\Delta\lambda_{HHH}=+1$
(dashed black line), $\Delta\lambda_{HHH}=-1$ (dotted black line), the
full set of ${\cal O}(\alpha_s^2\alpha^4)$ and ${\cal O}(\alpha^6)$
$\nu\bar\nu b\bar bb\bar b$ diagrams (magenta line), and the relevant
background processes is shown in Fig.~\ref{fig:six}.
In $ZHH$ production, positive (moderately negative) values of
$\Delta\lambda_{HHH}$ increase (decrease) the cross section. The
opposite is true for $\nu\bar\nu HH$ production. For
$\Delta\lambda_{HHH}=-1$, the cross section is strongly enhanced, while
for $\Delta\lambda_{HHH}=+1$ it is somewhat smaller than the SM cross section
over much of the $M_{HH}$ range. Only near the $HH$ threshold is the
cross section larger than in the SM. As in the $ZHH$ case, the
deviations from the SM are largely concentrated at small values of
$M_{HH}$. Comparison with Fig.~\ref{fig:two} shows that the cross
section for $\nu\bar\nu b\bar bb\bar b$ production is a few times larger
than that for $jjb\bar bb\bar b$ production at $\sqrt{s}=1$~TeV. 

When the full set of ${\cal O}(\alpha_s^2\alpha^4)$ and ${\cal
O}(\alpha^6)$ diagrams instead of the $\nu\bar\nu HH$ signal diagrams is
used to calculate the $\nu\bar\nu b\bar bb\bar b$ cross section, 
the resulting differential cross section is reduced by $10-20\%$, except
for the $HH$ threshold region, where it is enhanced. It is thus
relatively easy to confuse the effects of single- and non-resonant
Feynman diagrams and those of moderately positive anomalous Higgs
self-couplings. 

The $\nu\bar\nu b\bar bc\bar c$ and $\nu\bar\nu b\bar bjj$ backgrounds
are found to be at least one order of magnitude smaller than the
signal for the efficiencies used here. Due to the renormalization scale
uncertainty, the cross sections 
for these processes may vary by ${\cal O}(30-40\%)$. 
The $b\bar bb\bar b$ and $b\bar bjj$ backgrounds are largest for $M_{HH}$
close to the kinematic boundary, $\sqrt{s}$, and drop rapidly for
smaller values of $M_{HH}$. This is easily understood. In both cases, the 
invariant mass of the $b\bar bb\bar b$ system equals the center of mass
energy if 
the energy loss due to $b$-quark decays and jet mismeasurements are not taken
into account. While these backgrounds are substantial for
$M_{HH}>850$~GeV at a 1~TeV machine, they decrease rapidly with
increasing values of $\sqrt{s}$. 

The $b\bar bb\bar b$ and $b\bar
bjj$ backgrounds are concentrated at large values of $M_{HH}$, whereas
anomalous Higgs boson self-couplings mostly affect the cross section for
small values of the Higgs pair invariant mass. These backgrounds thus
have little or no effect on the sensitivity limits 
for $\lambda_{HHH}$. Since the region of large
Higgs pair invariant masses is largely insensitive to
$\lambda_{HHH}$, measuring the cross section in this region would
make it possible to normalize the cross section. Unfortunately, the
$e^+e^-\to\nu\bar\nu HH\to\nu\bar\nu b\bar bb\bar b$ 
cross section falls very quickly with increasing Higgs pair invariant
mass, in particular for higher center of mass energies, resulting in a
large statistical uncertainty in the large $M_{HH}$ region. Accurate
theoretical predictions of the 
SM $\nu\bar\nu b\bar bb\bar b$ cross section thus will be indispensable
for a measurement of the Higgs boson self-coupling in this final state.

\subsection{$m_H=140$~GeV}
\label{sec:fiveb}

The branching ratio for $H\to b\bar b$ drops rather quickly with
increasing Higgs boson mass 
and, for $m_H=140$~GeV, only about 1/3 of the Higgs bosons decay into
$b\bar b$. At the same time, the $H\to W^*W\to 4f$ branching ratio
increases to about 50\%. The $4f$ final state consists of four jets with
a probability of about 46\% while the $\ell\nu_\ell jj$ ($\ell=e,\,\mu$) final
state has a branching ratio of about 29\%. All other final states have a
combined branching ratio of 25\%. This suggests to consider the
$\nu\bar\nu b\bar b 4j$, $\nu\bar\nu b\bar b \ell\nu_\ell jj$,
$\nu\bar\nu 8j$ and $\nu\bar\nu \ell\nu_\ell 6j$ final states in
addition to $\nu\bar\nu b\bar bb\bar b$ production. Since the decay
$HH\to b\bar b 4j$ has the largest individual branching fraction of all
Higgs pair decays for 
$m_H=140$~GeV, I will consider $\nu\bar \nu b\bar b 4j$ production in
addition to the $\nu\bar \nu b\bar bb\bar b$ final state in this Section.

\begin{figure}[th!] 
\begin{center}
\includegraphics[width=11.1cm]{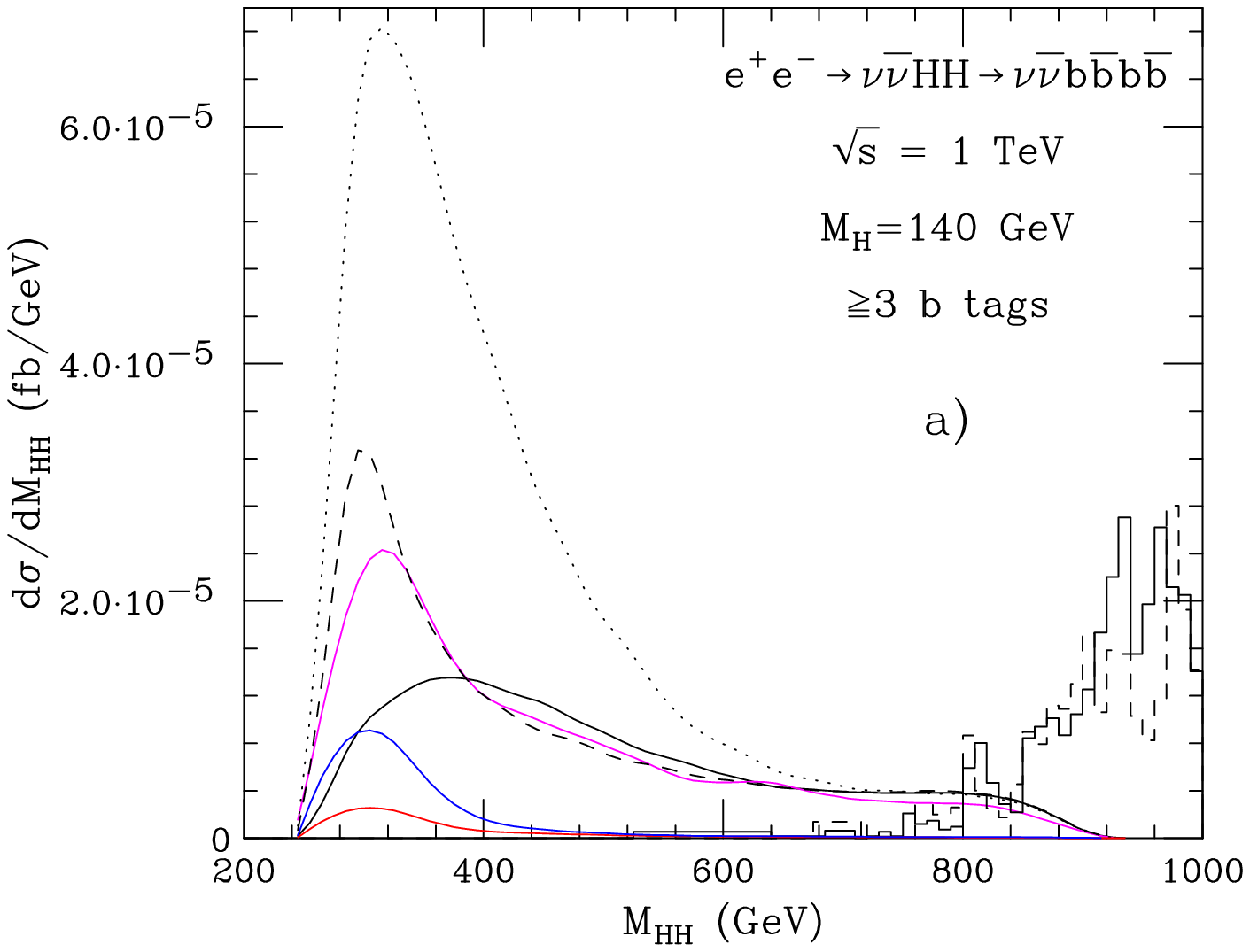} \\[3mm]
\includegraphics[width=11.1cm]{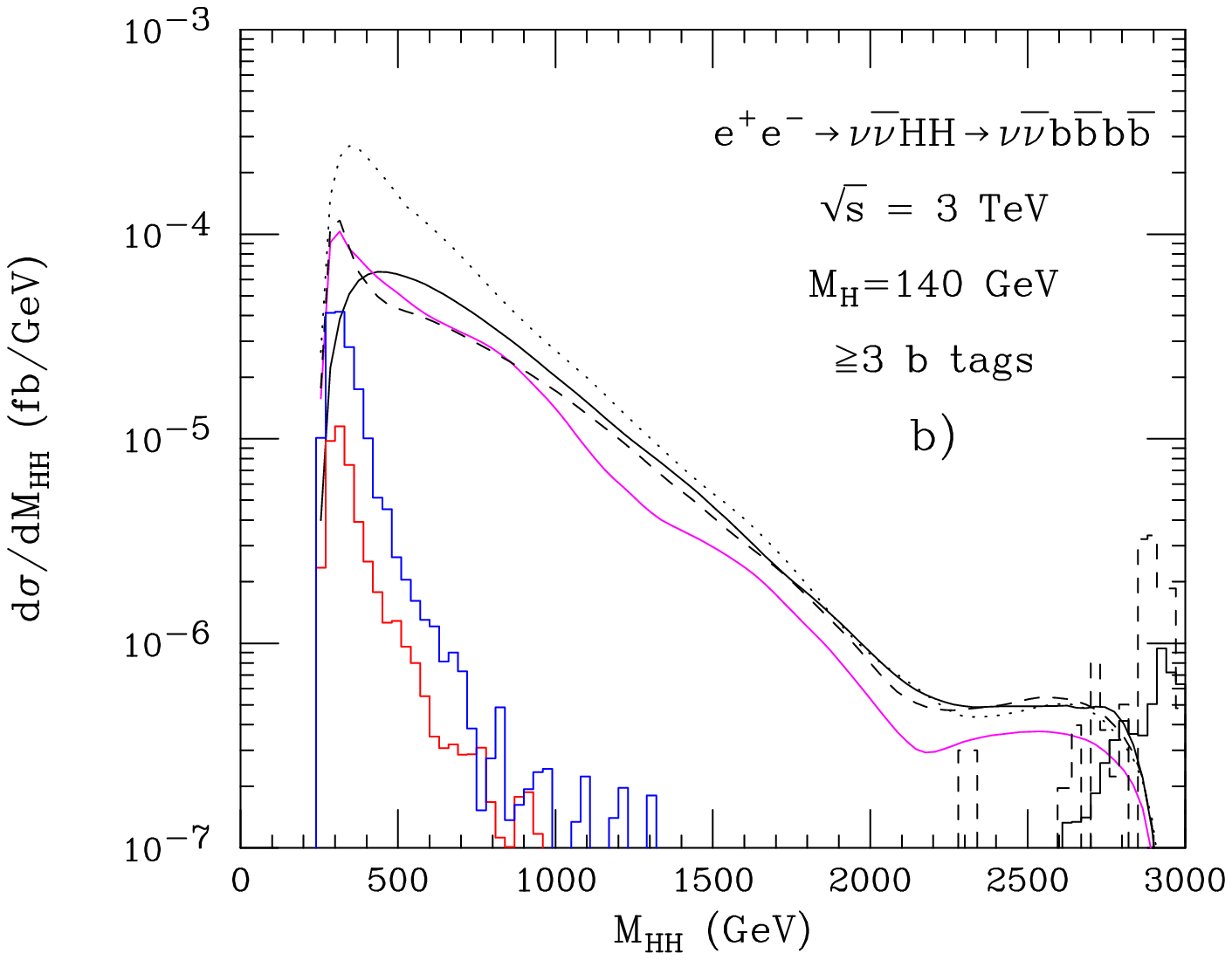}
\vspace*{2mm}
\caption[]{\label{fig:seven} 
The $e^+e^-\to\nu\bar\nu HH\to\nu\bar\nu b\bar bb\bar b$ cross section
with three or more $b$-tags as a function of
the Higgs pair invariant mass, $M_{HH}$, for $m_H=140$~GeV and a)
$\sqrt{s}=1$~TeV and b) $\sqrt{s}=3$~TeV. The black solid line is the
prediction of the SM signal cross section. The black dashed and dotted
lines correspond to the $\nu\bar\nu HH$ signal cross section for
$\Delta\lambda_{HHH}=+1$ and $\Delta\lambda_{HHH}=-1$, respectively. The
magenta line 
shows the SM cross section for $e^+e^-\to\nu\bar\nu b\bar bb\bar b$
including the full set of ${\cal O}(\alpha^6)$ 
and ${\cal O}(\alpha_s^2\alpha^4)$ Feynman diagrams. The blue
and red lines/histograms correspond to the $\nu\bar\nu b\bar bc\bar c$ and
$\nu\bar\nu b\bar bjj$ cross section. The solid and dashed histograms
represent the prediction of the $b\bar bb\bar b$ and $b\bar bjj$
backgrounds. The cuts  
imposed and the efficiencies used are summarized in Eqs.~(\ref{eq:cuts1}),
(\ref{eq:eff1}), (\ref{eq:mjj}), and~(\ref{eq:mbb1}). }
\vspace{-7mm}
\end{center}
\end{figure}

The results for $\nu\bar \nu b\bar bb\bar b$ production with
$m_H=140$~GeV are shown in Fig.~\ref{fig:seven}.
I impose the same cuts as in Sec.~\ref{sec:fivea}, except for the cut on
$m(b\bar b)$ which is replaced by that of Eq.~(\ref{eq:mbb1}). 
Taking the full set of ${\cal O}(\alpha^6)$ and ${\cal
O}(\alpha_s^2\alpha^4)$ Feynman diagrams into account in the calculation
considerably enhances the cross section (magenta line) in the $HH$
threshold region, while it reduces the $\nu\bar \nu b\bar bb\bar b$ rate
by about 
20\% for larger values of $M_{HH}$. The effect of the non-signal
diagrams contributing to $\nu\bar \nu b\bar bb\bar b$ production can be easily
confused with that of a positive anomalous Higgs self-coupling. The
$\nu\bar\nu b\bar bc\bar c$ background is substantial in the $HH$
threshold region, but falls very quickly with the Higgs pair invariant
mass, and becomes negligible for $M_{HH}>400$~GeV. The $\nu\bar\nu b\bar
bjj$ background is about a factor four smaller than that originating
from $\nu\bar\nu b\bar bc\bar c$ production for the efficiencies used
here. $b\bar bjj$ and $b\bar bb\bar b$ production constitute an
important background for 
$M_{HH}>800$~GeV ($M_{HH}>2.5$~TeV) at $\sqrt{s}=1$~TeV
($\sqrt{s}=3$~TeV). 

When calculating the cross section for $\nu\bar\nu HH\to\nu\bar\nu b\bar
b4j$ production care has to be taken. In addition to $H\to W^*W\to 4j$,
$H\to Z^*Z\to 4j$ also contributes, albeit with a much smaller branching
ratio ($B(H\to Z^*Z)\approx 10\%$ vs. $B(H\to W^*W)\approx 50\%$ for
$m_H=140$~GeV). To select $\nu\bar\nu HH\to\nu\bar\nu b\bar b4j$ events,
I require, in addition to the standard jet and missing transverse
momentum cuts of Eqs.~(\ref{eq:cuts1}), two tagged $b$-quarks
satisfying Eq.~(\ref{eq:mbb}) and four non-tagged jets with an invariant
mass 
\begin{equation}
\label{eq:fourj}
|m_H-m(4j)|<20~{\rm GeV}.
\end{equation}
One un-tagged jet pair has to be consistent with originating from a $W$
decay (see Eq.~(\ref{eq:mjj})). 

The main reducible backgrounds originate from $\nu\bar\nu 6j$ production
where two jets are mistagged as $b$-quarks, $\nu\bar\nu c\bar c4j$
production where both charm quarks are misidentified as $b$'s, and from
$e^+e^-\to b\bar 
b4j$ with the missing transverse momentum originating from jet
mismeasurements and the energy loss arising from
$b$-decays. Furthermore, non-resonant  $\nu\bar\nu b\bar b4j$ production
constitutes a potentially dangerous irreducible background. 

With the exception of $e^+e^-\to b\bar b4j$, these are processes with
eight particles in the final state. I have attempted to calculate 
the $e^+e^-\to\nu\bar\nu b\bar b4j$ cross section including the full set
of ${\cal 
O}(\alpha_s^4\alpha^4)$, ${\cal O}(\alpha_s^2\alpha^6)$ and ${\cal
O}(\alpha^8)$ diagrams using several publically available
programs. Generating the contributing Feynman diagrams via {\tt
MadGraph}~\cite{Maltoni:2002qb} resulted in well over $10^5$ Feynman
diagrams and took more than 200~hours of CPU time on a 3ghz Xeon
workstation. The evaluation of the cross section for this process using {\tt
MadEvent} would require computing resources significantly larger than
those available, and therefore was not attempted. Like {\tt MadEvent},
the current version of {\tt Sherpa}~\cite{Gleisberg:2008ta} is based on
a Feynman diagrammatic approach~\cite{Krauss:2001iv}, and thus is
expected to be too slow to calculate the cross section of $2\to 8$
processes. In the future, {\tt Sherpa} will use~\cite{krauss} {\tt
COMIX}~\cite{Gleisberg:2008fv}, a new matrix element generator which is
based on color dressed Berends-Giele recursive
relations~\cite{Duhr:2006iq}. This should allow for a much faster
evaluation of matrix elements, and thus make it possible to calculate
the cross sections of processes with $\geq 8$ particles in the final
state. 

For processes with many particles in the final state, {\tt
WHIZARD}~\cite{Kilian:2007gr} and {\tt
HELAC-PHEGAS}~\cite{Cafarella:2007pc} are potential alternatives to
{\tt MadEvent} and {\tt Sherpa}. {\tt WHIZARD} uses {\tt
O'Mega}~\cite{Moretti:2001zz} which implements an 
algorithm that collects all common sub-expressions in the sum over
Feynman diagrams contributing to a given scattering amplitude at tree
level. {\tt HELAC-PHEGAS} calculates matrix elements using recursive
Schwinger-Dyson equations. I have attempted to compute the $e^+e^-\to
\nu\bar\nu b\bar b4j$ cross section using {\tt WHIZARD} and {\tt
HELAC-PHEGAS}. The compilation of the {\tt WHIZARD} code was terminated
without a result after more than 
40~hours of CPU time on a 3ghz Xeon workstation. The {\tt
HELAC-PHEGAS} code compiled successfully, but failed to run with an
error in an underlying 
basic linux library. Another program which may be able to handle a
process such as $e^+e^-\to\nu\bar\nu b\bar b4j$ is {\tt
carlomat}~\cite{Kolodziej:2008py}, which, however, is not publically
available yet.

\begin{figure}[th!] 
\begin{center}
\includegraphics[width=11.6cm]{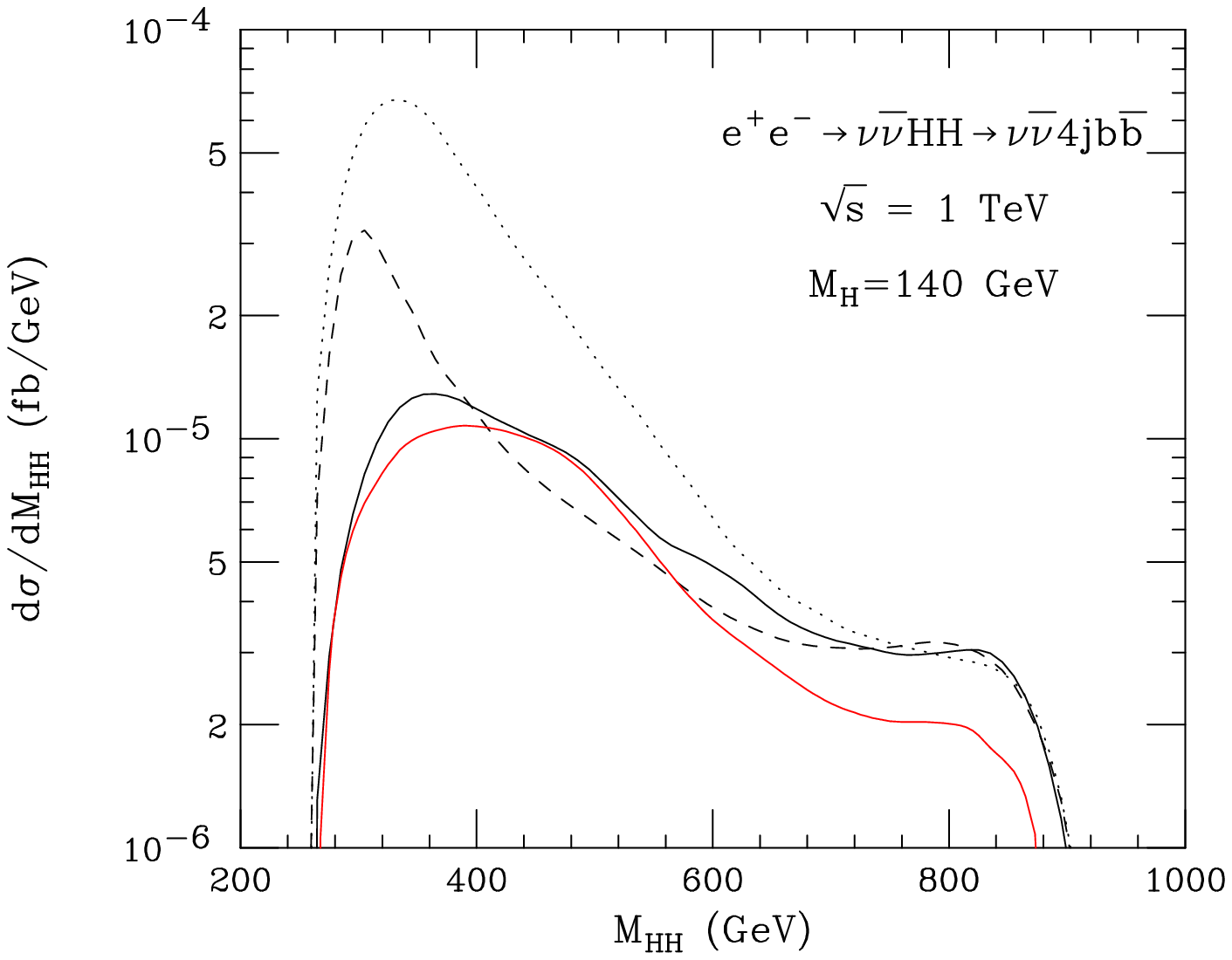} \\[3mm]
\includegraphics[width=11.6cm]{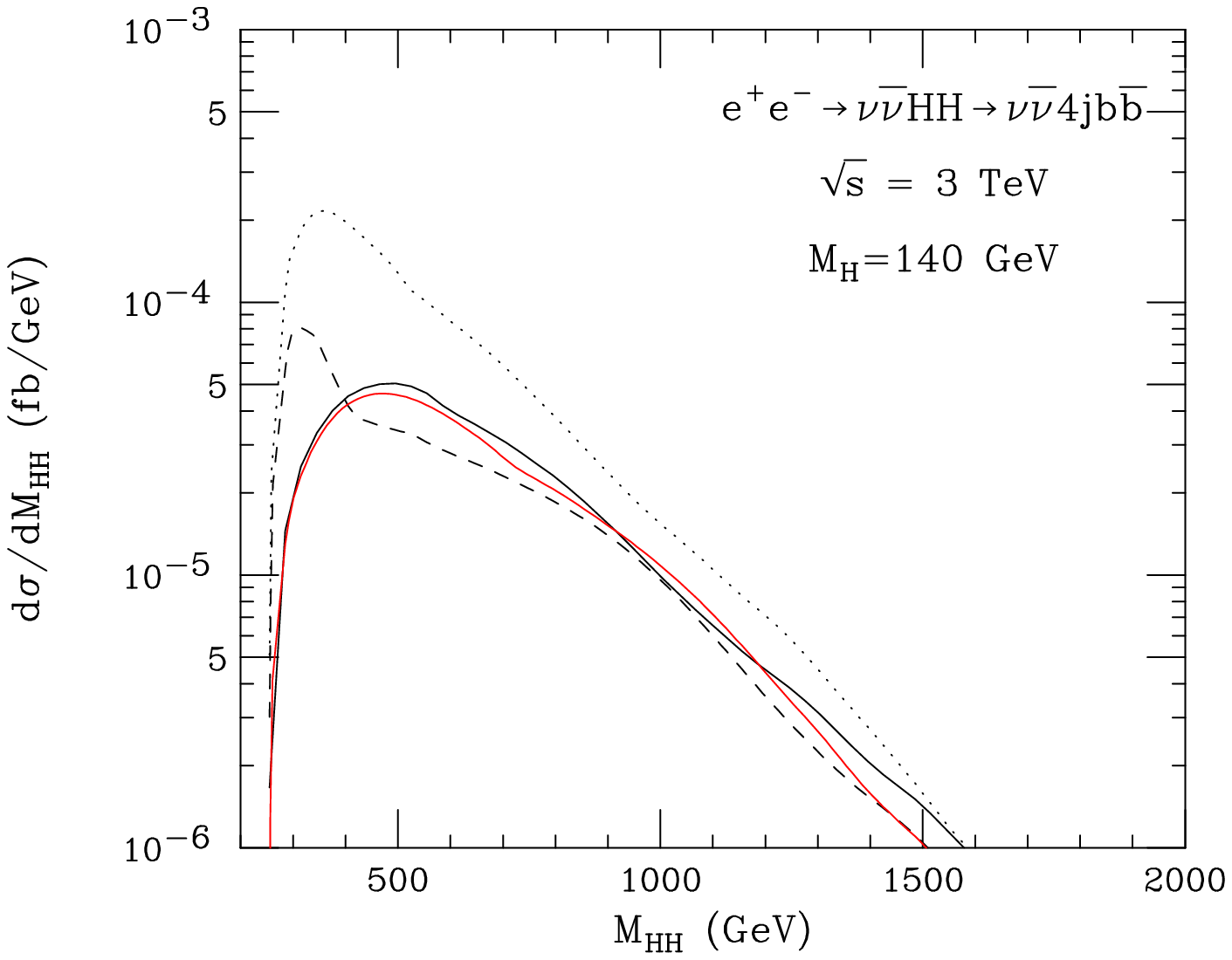}
\vspace*{2mm}
\caption[]{\label{fig:eight} 
The $e^+e^-\to\nu\bar\nu HH\to\nu\bar\nu b\bar b4j$ cross section as a
function of 
the Higgs pair invariant mass, $M_{HH}$, for $m_H=140$~GeV and a)
$\sqrt{s}=1$~TeV and b) $\sqrt{s}=3$~TeV. 
The black solid line is the
prediction of the SM signal cross section. The black dashed and dotted
lines correspond to the $\nu\bar\nu HH$ signal cross section for
$\Delta\lambda_{HHH}=+1$ and $\Delta\lambda_{HHH}=-1$, respectively. The
red line 
shows the SM cross section for $e^+e^-\to\nu\bar\nu Wjjb\bar b$, $W\to jj$,
including the full set of ${\cal O}(\alpha^7)$ 
and ${\cal O}(\alpha_s^2\alpha^5)$ Feynman diagrams. The cuts 
imposed and the efficiencies used are summarized in Eqs.~(\ref{eq:cuts1}),
(\ref{eq:eff1}), (\ref{eq:mjj}), (\ref{eq:mbb}), and~(\ref{eq:fourj}). } 
\vspace{-7mm}
\end{center}
\end{figure}

One can argue that a substantial portion of the contribution of the 
non-resonant ${\cal
O}(\alpha_s^2\alpha^6)$ and ${\cal O}(\alpha^8)$ diagrams\footnote{There
are no ${\cal O}(\alpha_s^4\alpha^4)$ diagrams contributing to the
process.} to the $\nu\bar\nu b\bar b4j$ cross section originates
from the off-shell $W^*\to jj$ jet pair. This suggests that calculating
the $e^+e^-\to\nu\bar\nu Wjjb\bar b$ cross section with $W\to jj$, or
the $\nu\bar\nu H4j$ cross section with $H\to b\bar b$, may be
sufficient for an estimate of how non-resonant diagrams affect the
$\nu\bar\nu b\bar b4j$ rate. Results based on a calculation of the process
$e^+e^-\to\nu\bar\nu Wjjb\bar b$ with $W\to jj$ (about 7,000 ${\cal
O}(\alpha_s^2\alpha^5)$ and ${\cal O}(\alpha^7)$ diagrams) are shown
in Fig.~\ref{fig:eight}.
The black solid line is the prediction of the SM signal cross section;
the black dashed and dotted 
lines correspond to the $\nu\bar\nu HH$ signal cross section for
$\Delta\lambda_{HHH}=+1$ and $\Delta\lambda_{HHH}=-1$, respectively. 
With the acceptance cuts imposed here, the $\nu\bar\nu b\bar b4j$ cross
section is approximately as large as that for $\nu\bar\nu b\bar bb\bar
b$ production. The
$e^+e^-\to\nu\bar\nu Wjjb\bar b$, $W\to jj$, cross section is shown by
the red line. While the non-resonant QCD and electroweak diagrams
included in $e^+e^-\to\nu\bar\nu Wjjb\bar b$, $W\to jj$, have a fairly
small effect at small values of $M_{HH}$, they are seen to reduce the
differential cross section by about 30\% in the large Higgs pair
invariant mass region. The non-resonant diagrams not included in 
$\nu\bar\nu Wjjb\bar b$, $W\to jj$ may well affect the cross section to
a similar degree. 

I have not attempted to compute the $\nu\bar\nu c\bar c4j$ and
$\nu\bar\nu 6j$ backgrounds, suspecting that one would run into problems
similar to those encountered for $e^+e^-\to\nu\bar\nu b\bar b4j$. I
expect that these backgrounds are small compared with the
signal, as in the case of the analogous $\nu\bar\nu b\bar bc\bar c$ and
$\nu\bar\nu 4j$ backgrounds in $\nu\bar\nu b\bar bb\bar b$ production.

The $e^+e^-\to b\bar b4j$ background (with about 15,000 ${\cal
O}(\alpha^6)$, ${\cal O}(\alpha_s^2\alpha^4)$ and ${\cal
O}(\alpha_s^4\alpha^2)$ Feynman diagrams
contributing) with the missing transverse
momentum originating from jet mismeasurements and the energy loss
arising from $b$-decays was found to be very small for the cuts
imposed. 

\subsection{$m_H=180$~GeV}
\label{sec:fivec}

For $m_H=180$~GeV, almost all Higgs bosons decay into a pair of
$W$-bosons ($B(H\to W^+W^-)\approx 93\%$). Subsequent $W$ decay then
leads to $HH\to\ell^\pm\nu_\ell 6j$ ($\ell=e,\,\mu$) with a branching ratio
of about 24\%, or $HH\to 8j$ with a branching fraction of $\approx
19\%$. Since the individual branching ratios for all other final states
are significantly smaller, I shall concentrate on $HH\to\ell^\pm\nu_\ell 6j$
and $HH\to 8j$ here.

If one of the four $W$-bosons decays leptonically, the final state
consists of one charged lepton, six jets and missing transverse momentum
which originates from the three neutrinos in the event. The main
backgrounds originate from single resonant and non-resonant
$e^+e^-\to\nu\bar\nu\ell^\pm\nu_\ell 6j$ diagrams and $W^\pm 6j$
production (with approximately 21,000 ${\cal O}(\alpha^3\alpha_s^4)$ 
Feynman diagrams contributing). 
In order to identify $\nu\bar\nu HH$ events in the $\ell^\pm p\llap/_T 6j$
final state, I require, in addition to the standard lepton, jet and
$p\llap/_T$ identification cuts of Eq.~(\ref{eq:cuts1}), three or more
jet pairs which satisfy 
Eq.~(\ref{eq:mjj}) with two of the jet pairs having an invariant mass in
the range 
\begin{equation}
\label{eq:4j}
160~{\rm GeV}<m([jj][jj])<200~{\rm GeV.}
\end{equation}
Since the invariant mass 
of the Higgs pair cannot easily be reconstructed, I consider the
distribution of the invariant mass of the six jets, $M_{6j}$, instead.

\begin{figure}[t!] 
\begin{center}
\includegraphics[width=15.5cm]{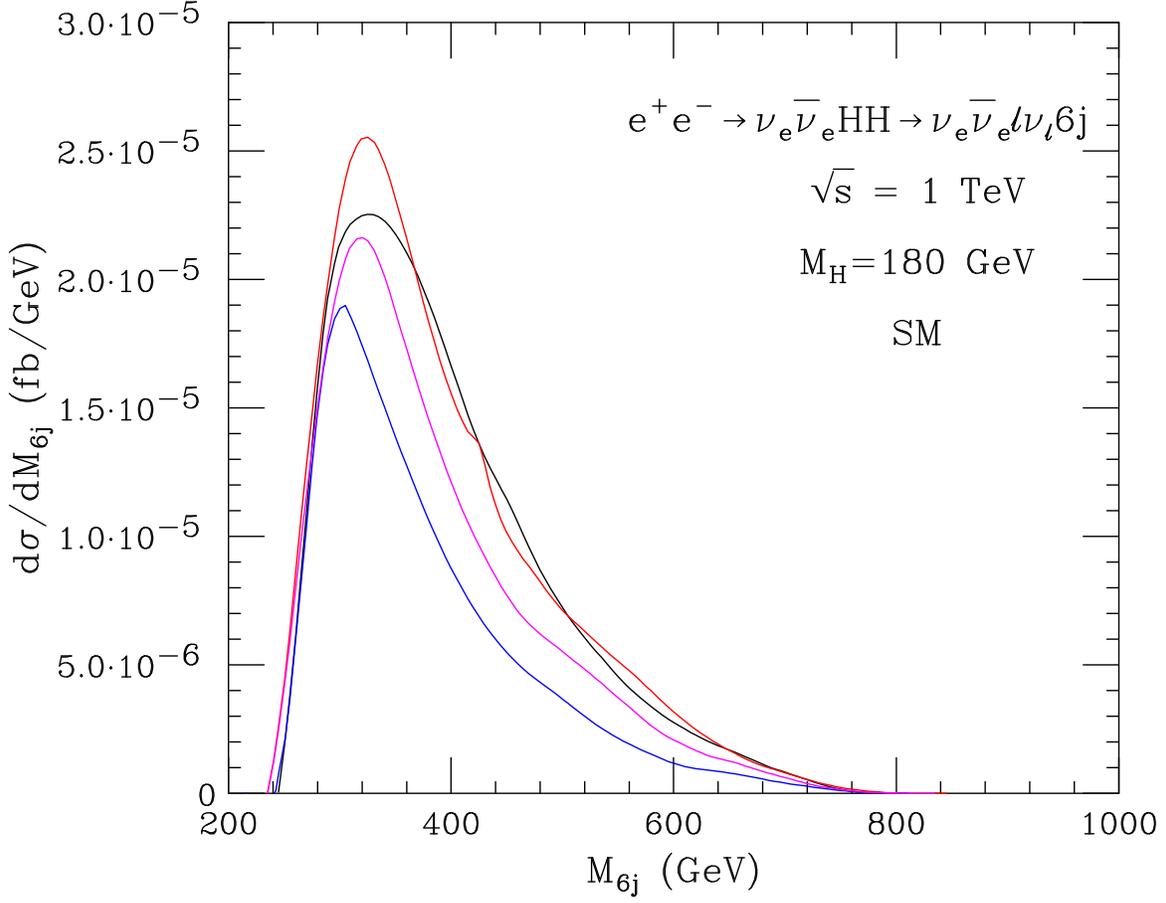}
\vspace*{2mm}
\caption[]{\label{fig:nine} 
The SM $e^+e^-\to\nu_e\bar\nu_e HH\to\nu_e\bar\nu_e\ell^\pm\nu_\ell 6j$
cross section as a function of
the $6j$ invariant mass, $M_{6j}$, for $\sqrt{s}=1$~TeV. The red
line shows the $e^+e^-\to \nu_e\bar\nu_e\ell^\pm\nu_\ell jjH$, $H\to
W^+W^-\to 4j$ cross section. The blue curve gives the
$e^+e^-\to\nu_e\bar\nu_e 4jH$, $H\to W^+W^-\to\ell\nu_\ell jj$ 
cross section. The magenta line gives an estimate of the
$e^+e^-\to\nu_e\bar\nu_e\ell^\pm\nu_\ell 6j$ cross section, obtained from
averaging the $e^+e^-\to \nu_e\bar\nu_e\ell^\pm\nu_\ell jjH$, $H\to
W^+W^-\to 4j$ and $e^+e^-\to\nu_e\bar\nu_e 4jH$, $H\to
W^+W^-\to\ell\nu_\ell jj$ differential cross sections. The cuts 
imposed and the efficiencies used are summarized in Eqs.~(\ref{eq:cuts1}),
(\ref{eq:eff1}), (\ref{eq:mjj}), and~(\ref{eq:4j}). }
\vspace{-7mm}
\end{center}
\end{figure}

Although a full calculation of the $2\to 10$ process
$e^+e^-\to\nu\bar\nu\ell^\pm\nu_\ell 6j$ is currently not feasible, it
is still possible to get an idea of how strongly non-resonant diagrams
affect the $6j$ invariant mass distribution by calculating the cross
sections of the processes $e^+e^-\to\nu\bar\nu\ell\nu_\ell jjH$ with
$H\to W^+W^-\to 4j$ (with approximately 1,300 ${\cal O}(\alpha^7)$ 
Feynman diagrams contributing), and $e^+e^-\to\nu\bar\nu 4jH$ with $H\to 
W^+W^-\to\ell\nu_\ell jj$ (with approximately 20,000 ${\cal
O}(\alpha^7)$ and ${\cal O}(\alpha^5\alpha_s^2)$ Feynman diagrams
contributing). In Fig.~\ref{fig:nine}, I show the SM
$e^+e^-\to\nu_e\bar\nu_eHH\to\nu_e\bar\nu_e\ell^\pm\nu_\ell 6j$ $M_{6j}$
differential cross section for the signal
at $\sqrt{s}=1$~TeV, together with the results for
$\nu_e\bar\nu_e\ell^\pm\nu_\ell jjH$, $H\to W^+W^-\to 4j$ production
(red line), and $e^+e^-\to\nu_e\bar\nu_e 4jH$, $H\to
W^+W^-\to\ell\nu_\ell jj$ (blue curve). 
Similar results are obtained for $\sqrt{s}=3$~TeV, and for
$e^+e^-\to\nu_l\bar\nu_lHH\to\nu_l\bar\nu_l\ell^\pm\nu_\ell 6j$ with
$l=\mu,\,\tau$. 

The non-resonant diagrams in $e^+e^-\to\nu_e\bar\nu_e\ell^\pm\nu_\ell
jjH$, $H\to 
W^+W^-\to 4j$ are found to significantly enhance the $\nu_e\bar\nu_e HH$
cross section near threshold. This is to be expected since the
$\ell\nu_\ell$ invariant mass cannot be constrained; the presence of
three neutrinos in the final state makes it impossible to use the $\ell
p\llap/_T$ transverse mass to reduce the background from non-resonant
diagrams. The non-resonant diagrams in $\nu_e\bar\nu_e 4jH$, $H\to
W^+W^-\to\ell\nu_\ell jj$ production, on the other hand, reduce the
cross section by a factor $1.5-2$ for the cuts imposed. 
An estimate of the effect of the full set of non-resonant diagrams may
be obtained by averaging the $\nu_e\bar\nu_e\ell^\pm\nu_\ell jjH$, $H\to
W^+W^-\to 4j$ and $\nu_e\bar\nu_e 4jH$, $H\to W^+W^-\to\ell\nu_\ell jj$
cross sections which is shown by the magenta line in
Fig.~\ref{fig:nine}. 

\begin{figure}[t!] 
\begin{center}
\includegraphics[width=15.5cm]{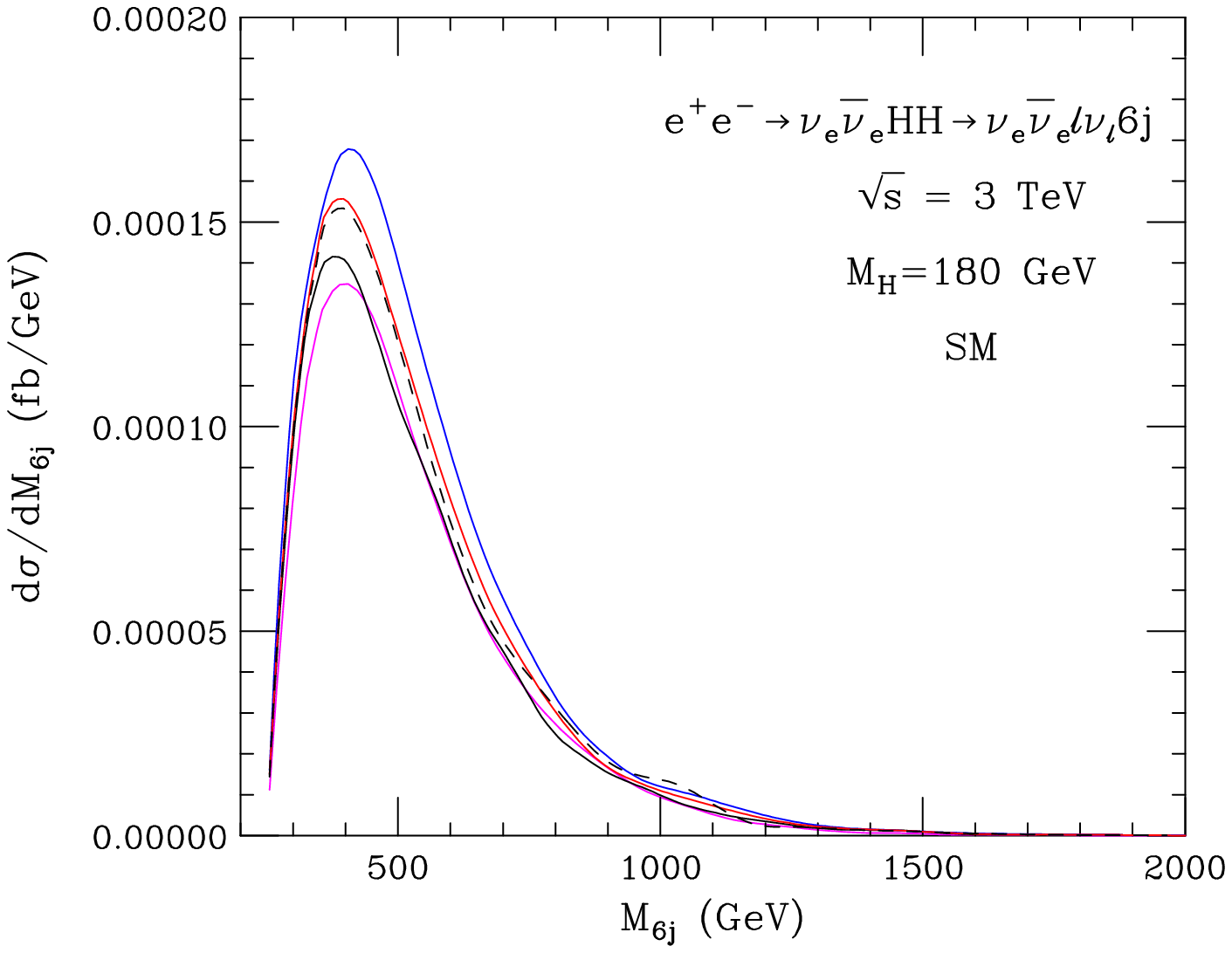}
\vspace*{2mm}
\caption[]{\label{fig:ten} 
The SM $e^+e^-\to\nu_e\bar\nu_e\ell^\pm\nu_\ell jjH$, $H\to WW\to 4j$
cross section as a function of
the $6j$ invariant mass, $M_{6j}$ for $\sqrt{s}=3$~TeV (dashed black
line). The black solid line shows the $\nu_e\bar\nu_e WjjH$,
$W\to\ell\nu_\ell$, $H\to WW\to 4j$ cross section. The blue curve gives the 
$\nu_e\bar\nu_e\ell^\pm\nu_\ell WH$, $W\to jj$, $H\to WW\to 4j$ 
cross section. The black dashed line gives an estimate of the
$e^+e^-\to\nu_e\bar\nu_e\ell^\pm\nu_\ell jjH$, $H\to WW\to 4j$ cross
section cross section from averaging the $\nu_e\bar\nu_e WjjH$,
$W\to\ell\nu_\ell$, $H\to WW\to 4j$ and $\nu_e\bar\nu_e\ell^\pm\nu_\ell
WH$, $W\to jj$, $H\to WW\to 4j$ differential cross sections. For
comparison, the magenta line shows the $\nu_e\bar\nu_e HH$,
$HH\to 4W\to\ell\nu_\ell 6j$ signal cross section. The cuts 
imposed and the efficiencies used are summarized in Eqs.~(\ref{eq:cuts1}),
(\ref{eq:eff1}), (\ref{eq:mjj}), and~(\ref{eq:4j}). }
\vspace{-7mm}
\end{center}
\end{figure}

The estimate obviously ignores a large number of non-resonant Feynman
diagrams. As demonstrated by the blue line in Fig.~\ref{fig:nine},
non-resonant Feynman diagrams may significantly affect the
$\nu_e\bar\nu_e HH$ cross section, and one may worry whether the
averaging procedure employed here does yield credible results. To
justify the averaging procedure, I compare in Fig.~\ref{fig:ten} the SM
$e^+e^- \to\nu_e\bar\nu_e\ell^\pm\nu_\ell jjH$, $H\to WW\to 4j$ $M_{6j}$
distribution for $m_H=180$~GeV and $\sqrt{s}=3$~TeV (black dashed line)
with the result obtained from averaging the
$\nu_e\bar\nu_e\ell^\pm\nu_\ell WH$, $W\to jj$, $H\to WW\to 4j$  (blue
line) and $\nu_e\bar\nu_e WjjH$, $W\to\ell\nu_\ell$ $H\to WW\to 4j$
(black solid line) cross sections (red line). 
For comparison, the magenta line shows the $\nu_e\bar\nu_e HH$,
$HH\to 4W\to\ell\nu_\ell 6j$ signal cross section. The red and black
dashed lines agree within a few percent, lending credibility to the
averaging procedure used in Fig.~\ref{fig:nine}. 

\begin{figure}[th!] 
\begin{center}
\includegraphics[width=11.5cm]{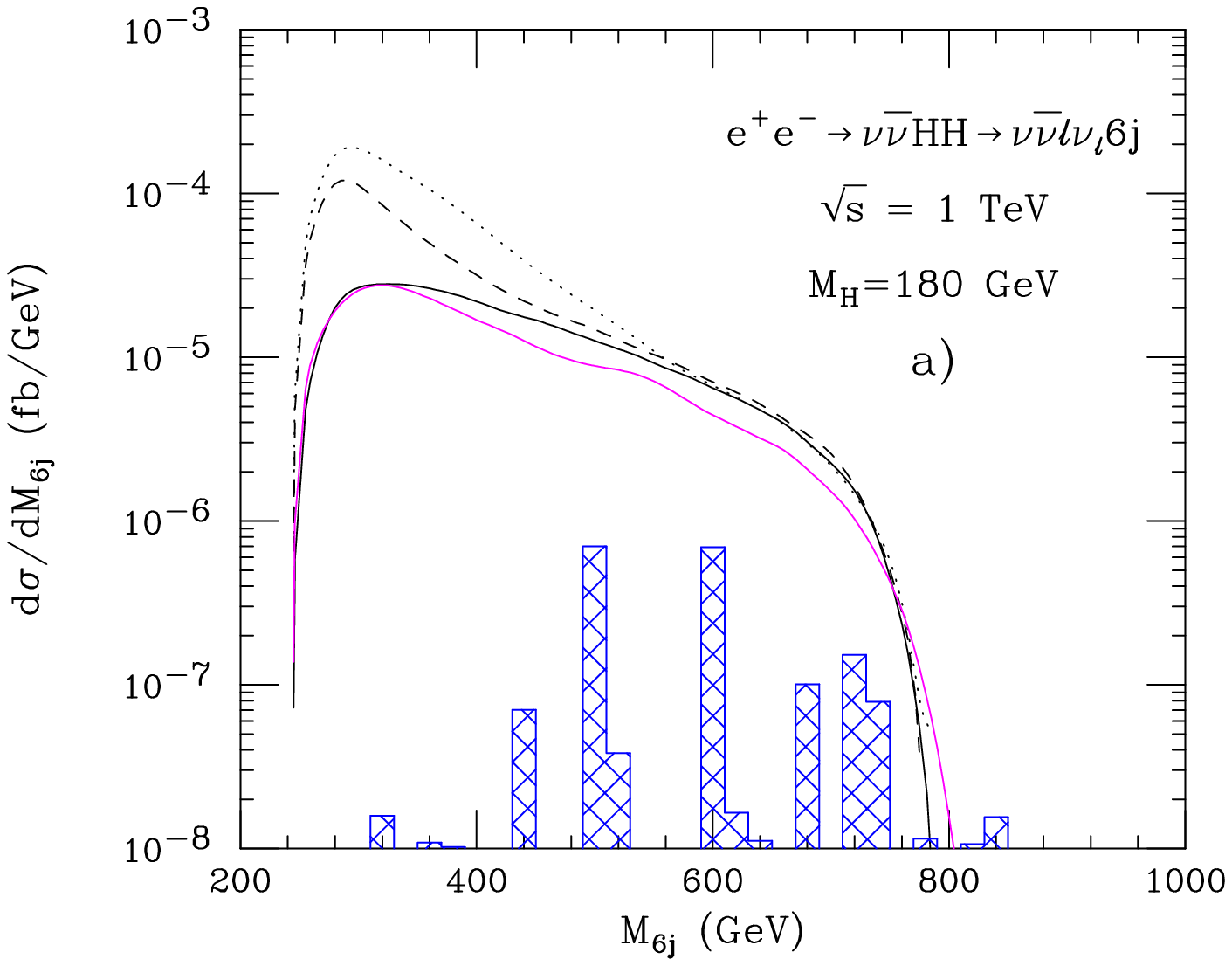} \\[3mm]
\includegraphics[width=11.5cm]{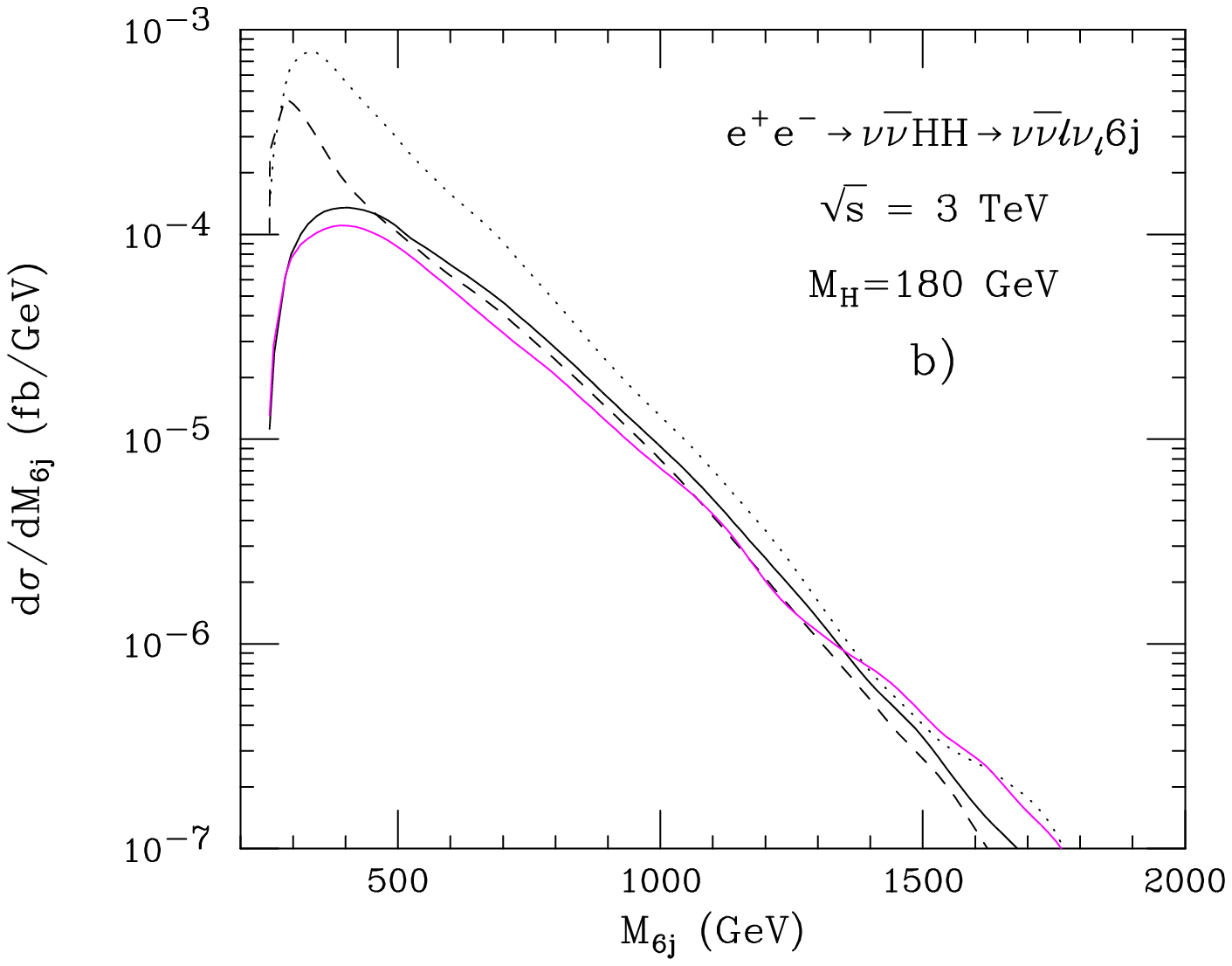}
\vspace*{2mm}
\caption[]{\label{fig:eleven} 
The $e^+e^-\to\nu\bar\nu HH\to\nu\bar\nu\ell\nu_\ell 6j$  cross section as a
function of 
the $6j$ invariant mass, $M_{6j}$, for $m_H=180$~GeV and a)
$\sqrt{s}=1$~TeV and b) $\sqrt{s}=3$~TeV. 
The black solid line is the
prediction of the SM signal cross section. The black dashed and dotted
lines correspond to the $\nu\bar\nu HH$ signal cross section for
$\Delta\lambda_{HHH}=+1$ and $\Delta\lambda_{HHH}=-1$, respectively. The
magenta line shows the estimated SM cross section for
$e^+e^-\to\nu\bar\nu\ell\nu_\ell 6j$. The blue hatched histogram in part
a) shows the SM $e^+e^-\to W6j$ background. The cuts 
imposed and the efficiencies used are summarized in Eqs.~(\ref{eq:cuts1}),
(\ref{eq:eff1}), (\ref{eq:mjj}), and~(\ref{eq:4j}). }
\vspace{-7mm}
\end{center}
\end{figure}

Adopting the averaging procedure introduced above, I show in
Fig.~\ref{fig:eleven} the $6j$
invariant mass distribution for the SM $e^+e^-\to\nu\bar\nu HH\to
\nu\bar\nu\ell\nu_\ell 6j$ signal (solid black line),
$\Delta\lambda_{HHH}=+1$ (dashed black line), and
$\Delta\lambda_{HHH}=-1$ (dotted black line) for $\sqrt{s}=1$~TeV and
3~TeV, together with the 
estimated $\nu\bar\nu\ell\nu_\ell 6j$ cross section including
non-resonant diagrams (magenta lines).
The non-resonant diagrams are seen to somewhat reduce the cross section
over most of the $6j$ invariant mass range. Only for large values of
$M_{6j}$ do they increase the differential cross section. The blue
hatched histogram in Fig.~\ref{fig:eleven}a shows the $W6j$
background. For $\sqrt{s}=3$~TeV the $W6j$ cross section is too small to
show up for the range of cross sections displayed. Clearly, the $e^+e^-\to
W6j$ background is negligible at both center of mass energies.

If all four $W$ bosons in $HH\to 4W$ decay hadronically, the final state
consists of eight jets and missing transverse momentum. To identify
$\nu\bar\nu HH$ events in the $p\llap/_T8j$ final state, I require, in
addition to the standard $p\llap/_T$ and jet identification cuts of
Eq.~(\ref{eq:cuts1}), four jet pairs which satisfy Eq.~(\ref{eq:mjj}).
Furthermore, the jet pairs are required to form two groups of four jet
systems which satisfy Eq.~(\ref{eq:4j}). The four jets have to consist
of two jet pairs with each $jj$ system fulfilling Eq.~(\ref{eq:mjj}). 

\begin{figure}[th!] 
\begin{center}
\includegraphics[width=11.7cm]{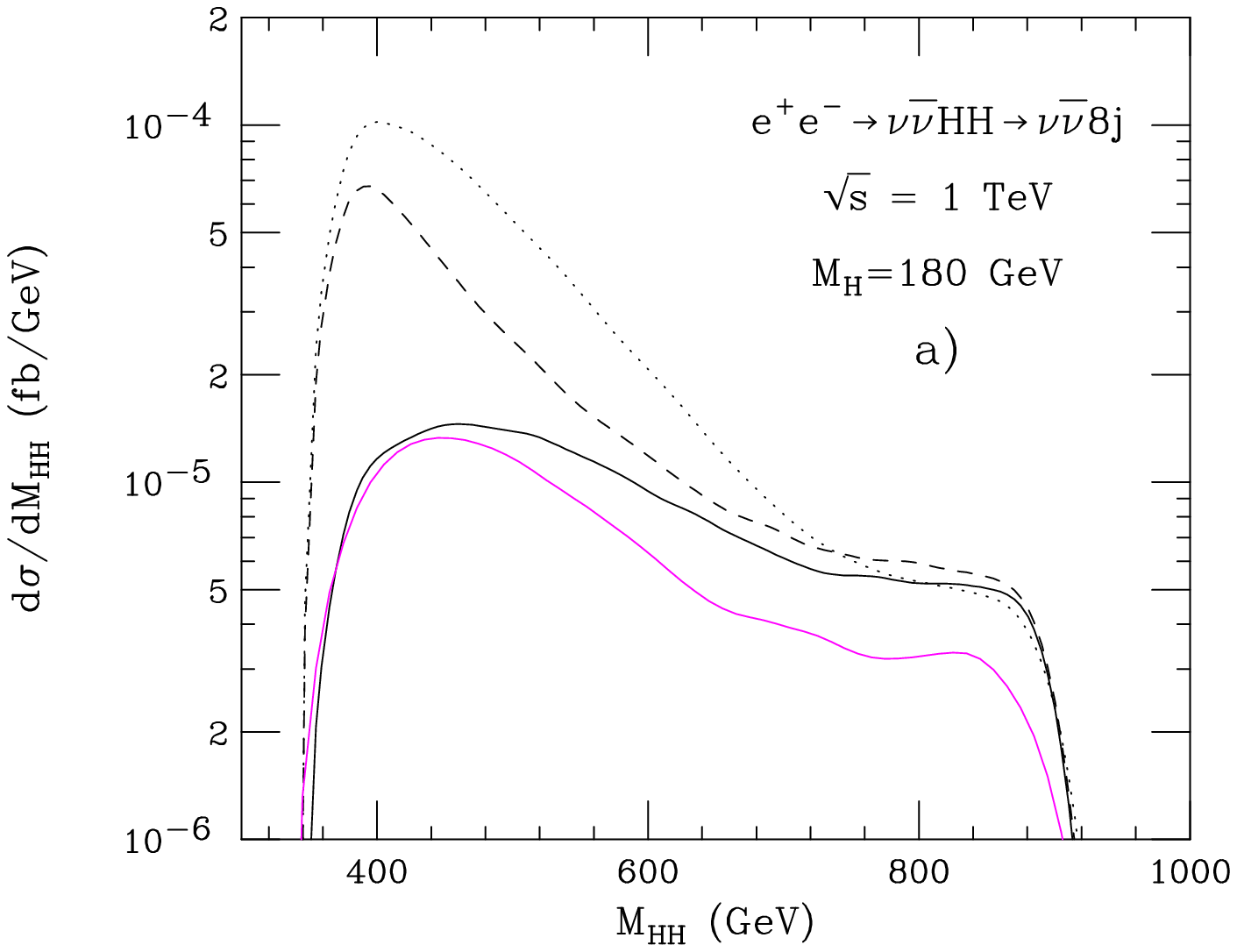} \\[3mm]
\includegraphics[width=11.7cm]{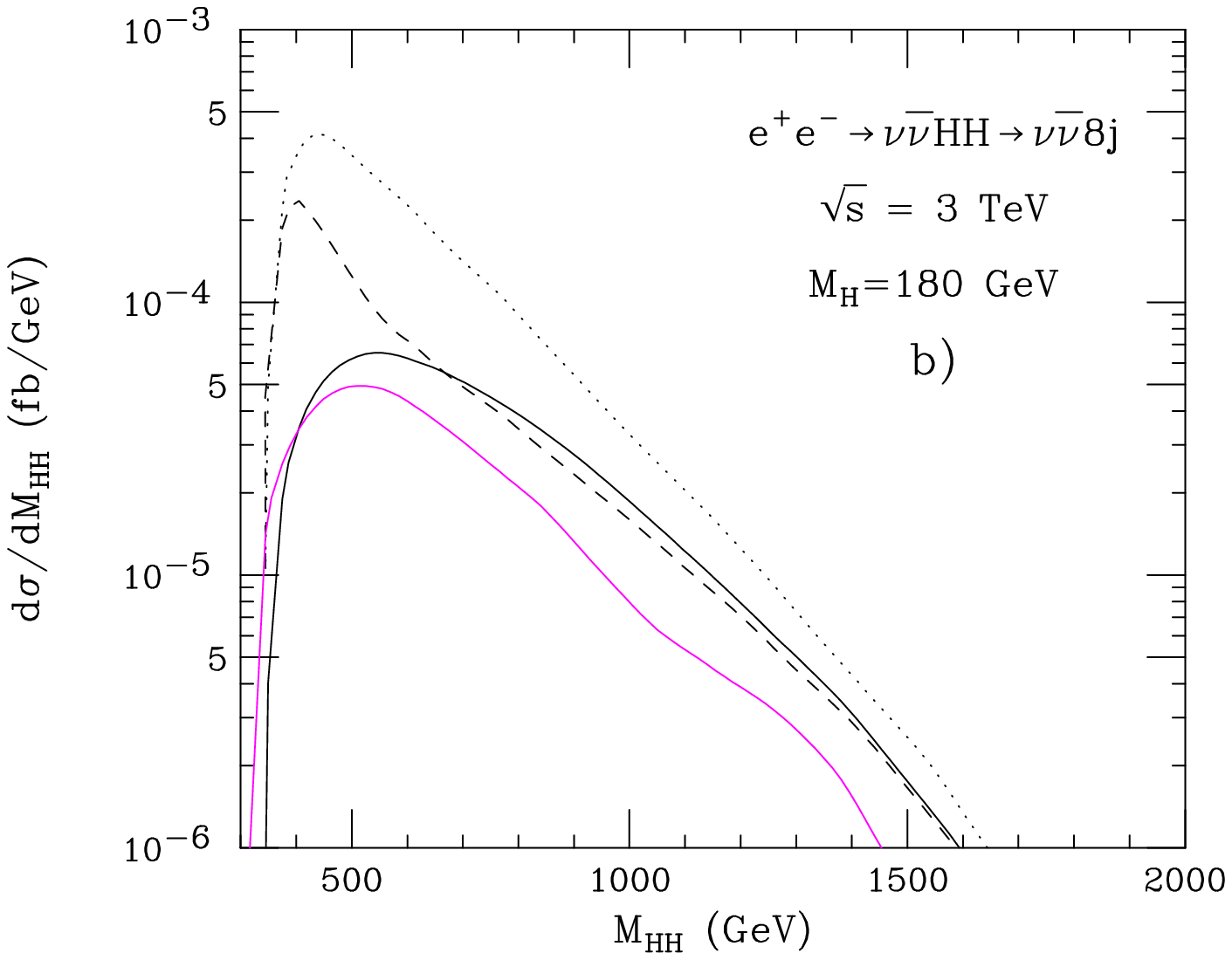}
\vspace*{2mm}
\caption[]{\label{fig:twelve} 
The $e^+e^-\to\nu\bar\nu HH\to\nu\bar\nu 8j$  cross section as a
function of the Higgs pair invariant mass, for $m_H=180$~GeV and a)
$\sqrt{s}=1$~TeV and b) $\sqrt{s}=3$~TeV. 
The black solid line is the
prediction of the SM signal cross section. The black dashed and dotted
lines correspond to the $\nu\bar\nu HH$ signal cross section for
$\Delta\lambda_{HHH}=+1$ and $\Delta\lambda_{HHH}=-1$, respectively. The
magenta line shows the estimated SM cross section for
$e^+e^-\to\nu\bar\nu 8j$ obtained from the $\nu\bar\nu 4jH$, $H\to WW\to
4j$ cross section. The cuts 
imposed and the efficiencies used are summarized in Eqs.~(\ref{eq:cuts1}),
(\ref{eq:eff1}), (\ref{eq:mjj}), and~(\ref{eq:4j}). }
\vspace{-7mm}
\end{center}
\end{figure}

The main background to the $p\llap/_T8j$ final state originates from
non-resonant diagrams which I estimate by calculating the cross section
for $e^+e^-\to\nu\bar\nu 4jH$ with $H\to WW\to 4j$ and employing the
same averaging procedure as for the $\ell p\llap/_T 6j$ final
state. The background from $e^+e^-\to 8j$ with the missing transverse
momentum originating from jet mismeasurements is expected to be small
except for Higgs pair invariant masses close to $\sqrt{s}$.
The results for the Higgs pair invariant mass distribution with
$\sqrt{s}=1$~TeV and 3~TeV are shown in Fig.~\ref{fig:twelve}.
The non-resonant diagrams included in $e^+e^-\to\nu\bar\nu 4jH$, $H\to
WW\to 4j$, are seen to substantially reduce the cross
section away from the threshold (magenta line). There is no guarantee
that the averaging procedure used approximates the $\nu\bar\nu 8j$ cross
section including the full set of non-resonant Feynman diagrams with a
similar accuracy as that observed for the $\nu\bar\nu\ell\nu_\ell 6j$
final state. The non-resonant diagrams ignored in $e^+e^-\to\nu\bar\nu
4jH$, $H\to WW\to 4j$ may increase, or further decrease, the cross
section. The main conclusion drawn from Fig.~\ref{fig:twelve},
therefore, is that non-resonant Feynman diagrams may substantially
affect the $\nu\bar\nu 8j$ cross section, and alter the shape of the
$M_{HH}$ distribution. The effect of non-resonant Feynman diagrams can
of course be reduced by imposing a more stringent cut on the $4j$
invariant mass than that used here (see Eq.~(\ref{eq:4j})). Whether this
will be possible largely depends on the resolution of the hadronic
calorimeter which can be achieved in an ILC/CLIC experiment.

\subsection{Compilation of cross sections}
\label{sec:fived}

Before I derive sensitivity limits for the Higgs boson self-coupling, I
present, in Table~\ref{tab:two}, integrated signal and background cross
sections for $e^+e^- \to\nu\bar\nu HH$ and the Higgs boson masses and
final states discussed in Secs.~\ref{sec:fivea} --~\ref{sec:fivec}.
\begin{table}
\caption[]{\label{tab:two}
Cross sections in fb for $e^+e^-\to\nu\bar\nu b\bar bb\bar b$ for
$m_H=120$~GeV and 
$m_H=140$~GeV with three or more $b$-tags, $e^+e^-\to\nu\bar\nu 4jb\bar
b$ for $m_H=140$~GeV, and
$e^+e^-\to\nu\bar\nu\ell\nu_\ell 6j$ and $e^+e^-\to\nu\bar\nu 8j$ for
$m_H=180$~GeV. Results are shown for $\sqrt{s}=1$~TeV and 3~TeV, the SM
signal, the full set of Feynman diagrams (labeled ``all''), and the
reducible backgrounds (labeled ``bgd''). The reducible background does
not include 
contributions where the missing transverse momentum originates only from
the energy loss of $b$-quarks and/or jet mismeasurements. The cuts and
efficiencies used are listed in Eqs.~(\ref{eq:cuts1}), (\ref{eq:eff1}),
(\ref{eq:mjj}), (\ref{eq:mbb}), (\ref{eq:fourj}), and (\ref{eq:4j}).} 
\vspace{2mm}
\begin{tabular}{c|ccc}
\multicolumn{4}{c}{$e^+e^-\to\nu\bar\nu b\bar bb\bar b$}\\
\hline
\phantom{j} & signal & all & bgd \\
\hline
$\sqrt{s}=1$~TeV, $m_H=120$~GeV & 0.026 & 0.024 & 0.003\\
$\sqrt{s}=1$~TeV, $m_H=140$~GeV & 0.0044 & 0.0052 & 0.0014\\
\hline
$\sqrt{s}=3$~TeV, $m_H=120$~GeV & 0.20 & 0.16 & 0.01\\
$\sqrt{s}=3$~TeV, $m_H=140$~GeV & 0.040 & 0.037 & 0.006\\
\hline
\multicolumn{4}{c}{$e^+e^-\to\nu\bar\nu 4jb\bar b$}\\
\hline
\phantom{j} & signal & all & bgd \\
\hline
$\sqrt{s}=1$~TeV, $m_H=140$~GeV & 0.0039 & 0.0032 & -- \\
$\sqrt{s}=3$~TeV, $m_H=140$~GeV & 0.025 & 0.023 & -- \\
\hline
\multicolumn{4}{c}{$e^+e^-\to\nu\bar\nu\ell\nu_\ell 6j$}\\
\hline
\phantom{j} & signal & all & bgd \\
\hline
$\sqrt{s}=1$~TeV, $m_H=180$~GeV & 0.0065 & 0.0054 & $4.0\times 10^{-5}$ \\
$\sqrt{s}=3$~TeV, $m_H=180$~GeV & 0.050 & 0.041 & $<1\times 10^{-6}$ \\
\hline
\multicolumn{4}{c}{$e^+e^-\to\nu\bar\nu 8j$}\\
\hline
\phantom{j} & signal & all & bgd \\
\hline
$\sqrt{s}=1$~TeV, $m_H=180$~GeV & 0.0047 & 0.0036 & -- \\
$\sqrt{s}=3$~TeV, $m_H=180$~GeV & 0.032 & 0.022 & --
\end{tabular}
\end{table}
Table~\ref{tab:two} shows that, in contrast to $ZHH$ production, the
background is always small in $e^+e^-\to\nu\bar\nu HH$. Furthermore, the
$\nu\bar\nu HH\to\nu\bar\nu b\bar bb\bar b$ cross section is
considerably larger 
than that for $ZHH\to jjb\bar bb\bar b$ for $m_H=120$~GeV and
$m_H=140$~GeV. Non-resonant diagrams have a moderate effect except for
$\nu\bar\nu HH\to\nu\bar\nu b\bar bb\bar b$ production and
$m_H=140$~GeV, and for 
$e^+e^-\to\nu\bar\nu 8j$. For $\nu\bar\nu b\bar bb\bar b$ production,
QCD diagrams 
contribute at ${\cal O}(\alpha_s^2\alpha^4)$. Since these diagrams
contribute only $10-30\%$ of the cross section over most of the $M_{HH}$
range, I estimate the renormalization scale uncertainty of the
$\nu\bar\nu HH\to\nu\bar\nu b\bar bb\bar b$ cross section to be no more
than 10\%.  

The main uncertainty of the $e^+e^-\to\nu\bar\nu 4jb\bar b$,
$e^+e^-\to\nu\bar\nu\ell\nu_\ell 6j$ and $e^+e^-\to\nu\bar\nu 8j$ cross
sections originates from the unknown effect of non-resonant
diagrams. Unfortunately, calculations of these processes including the
full set of contributing Feynman diagrams is currently beyond what
automated matrix element based programs can handle. I have presented
results based on approximations which include subsets of non-resonant
diagrams, and argued that these should account for most of the effects
of non-resonant diagrams. Nevertheless, an uncertainty of ${\cal
O}(20\%)$ (${\cal O}(50\%)$) remains for the $e^+e^-\to\nu\bar\nu 4jb\bar
b$ and $e^+e^-\to\nu\bar\nu\ell\nu_\ell 6j$ ($e^+e^-\to\nu\bar\nu 8j$)
cross section. 

\section{Sensitivity limits for $\Delta\lambda_{HHH}$}
\label{sec:six}

I now present quantitative sensitivity limits for the Higgs boson
self-coupling for $e^+e^-\to ZHH$ and $e^+e^-\to\nu\bar\nu HH$, and the
final states discussed in 
Secs.~\ref{sec:four} and~\ref{sec:five}. Limits are derived from the
$M_{HH}$ distribution except for $e^+e^-\to\nu\bar\nu\ell\nu_\ell 6j$
where the $6j$ invariant mass distribution is analyzed. Results are
presented for integrated luminosities of 0.5~ab$^{-1}$, 1~ab$^{-1}$, and
2~ab$^{-1}$ at $\sqrt{s}=0.5$~TeV and 1~TeV (ILC), and 1~ab$^{-1}$,
2~ab$^{-1}$, and 3~ab$^{-1}$ at $\sqrt{s}=3$~TeV (CLIC). According to
the current ILC (CLIC) design
parameters~\cite{Brau:2007zza,Braun:2008zz}, an integrated luminosity of
1~ab$^{-1}$ (3~ab$^{-1}$) corresponds to 5~years of running. For $ZHH$
production, I analyze the $jjb\bar bb\bar b$ final state with $\geq 3$ and
4~$b$-tags. For $\sqrt{s}=0.5$~TeV, bounds are calculated for the sets
of efficiencies listed in Eqs.~(\ref{eq:eff1}) and~(\ref{eq:eff2}),
otherwise I only use the efficiencies of Eq.~(\ref{eq:eff1}). To derive
limits, I use the cross sections obtained including non-resonant Feynman
diagrams. 

As the statistical tool of choice
I adopt a log likelihood test. The expression for the log-likelihood
function is 
\begin{eqnarray}
\nonumber
-2\log L & = &
-2\left[\sum_i\left(-f_SS_i-f_BB_i+n_{0i}\log(f_SS_i 
                    +f_BB_i) - \log(n_{0i}!)\right)\right]
\\ 
& & + {(f_S-1)^2\over(\Delta f_S)^2}
    + {(f_B-1)^2\over(\Delta f_B)^2} \, .
\end{eqnarray}
The sum extends over the number of bins, $S_i$ and $B_i$ are the
number of signal and background events in the $i$th bin, and $n_{0i}$
is the number of reference (eg. SM) events in the $i$th bin.  The
uncertainties on the 
signal and background normalizations are taken into account via two
multiplicative factors, $f_S$ and $f_B$, which are allowed to vary but
are constrained within the relative uncertainties of the signal and
background cross sections, $\Delta f_S$ and $\Delta f_B$,
respectively. 

In order to simplify the analysis, I assume a common uncertainty for
signal and background, $f_S=f_B=f$, in the
following. This can be justified by noting that either the signal is
considerably larger than the background, or vice versa. In the first
case, I use the numerical value of $f_S$, in the second, $f_B$ is
used. The uncertainties, $\Delta f$, are determined individually from
the renormalization scale uncertainty (if QCD diagrams contribute) or the
approximation used to calculate the cross section. These
uncertainties have been discussed in Secs.~\ref{sec:four}
and~\ref{sec:five}. In all other cases, a generic theory uncertainty of
10\% is assumed to account for unknown higher order electroweak
corrections. The values of $\Delta f$ used in the following analysis are
collected in Table~\ref{tab:three}.
\begin{table}
\caption[]{\label{tab:three}
Theoretical uncertainties, $\Delta f$, used in the statistical analysis.} 
\vspace{2mm}
\begin{tabular}{c|cc|c|cc}
\multicolumn{6}{c}{$e^+e^-\to jjb\bar bb\bar b$, $\epsilon_b=0.9$,
$P_{c\to b}=0.1$, $P_{j\to b}=0.005$} \\
\hline
$m_H=120$~GeV & 4 $b$-tags & $\geq 3$ $b$-tags & $m_H=140$~GeV & 4 $b$-tags
& $\geq 3$ $b$-tags\\ 
\hline
$\sqrt{s}=0.5$~TeV & 10\% & 100\% & 
$\sqrt{s}=0.5$~TeV & 40\% & 100\% \\
$\sqrt{s}=1$~TeV & 10\% & 10\% & 
$\sqrt{s}=1$~TeV & 10\% & 30\% \\
\hline
\multicolumn{6}{c}{$e^+e^-\to jjb\bar bb\bar b$, $\epsilon_b=0.8$,
$P_{c\to b}=0.02$, $P_{j\to b}=0.001$} \\
\hline
 $m_H=120$~GeV & 4 $b$-tags & $\geq 3$ $b$-tags & $m_H=140$~GeV & 4 $b$-tags
& $\geq 3$ $b$-tags\\ 
\hline
$\sqrt{s}=0.5$~TeV & 10\% & 40\% & 
$\sqrt{s}=0.5$~TeV & 10\% & 100\% \\
\hline
\multicolumn{6}{c}{$e^+e^-\to\nu\bar\nu b\bar bb\bar b$} \\
\hline
 $m_H$ & $\sqrt{s}=1$~TeV  & $\sqrt{s}=3$~TeV  & $m_H$ &
$\sqrt{s}=1$~TeV  & $\sqrt{s}=3$~TeV \\ 
\hline
120~GeV & 10\% & 10\% & 140~GeV & 10\% & 10\% \\
\end{tabular}
\begin{tabular}{cc|cc|cc}
\multicolumn{2}{c|}{$e^+e^-\to\nu\bar\nu 4jb\bar b$, $m_H=140$~GeV} &
\multicolumn{2}{c|}{$e^+e^-\to\nu\bar\nu\ell\nu_\ell 6j$, $m_H=180$~GeV}
& \multicolumn{2}{c}{$e^+e^-\to\nu\bar\nu 8j$, $m_H=180$~GeV} \\
\hline
$\sqrt{s}=1$~TeV  & $\sqrt{s}=3$~TeV  & $\sqrt{s}=1$~TeV  &
$\sqrt{s}=3$~TeV  & $\sqrt{s}=1$~TeV  & $\sqrt{s}=3$~TeV \\ 
\hline
20\% & 20\% & 20\% & 20\% & 50\% & 50\% 
\end{tabular}
\end{table}

The rather large uncertainties listed for $\sqrt{s}=0.5$~TeV, 
$m_H=140$~GeV, and/or $\geq 3$ $b$-tags originate from the large
renormalization uncertainty of the (reducible) background and could be
reduced by a NLO QCD calculation of $e^+e^-\to b\bar bcjjj$ and
$e^+e^-\to b\bar bc\bar cjj$. Such calculations are beyond the current
state of the art of one-loop calculations. Likewise, the uncertainties
listed for $e^+e^-\to\nu\bar\nu 4jb\bar b$,
$e^+e^-\to\nu\bar\nu\ell\nu_\ell 6j$ and $e^+e^-\to\nu\bar\nu 8j$ could
potentially be reduced by performing a full matrix element based
tree level calculation of these processes. Further advances in automated
tree level programs may make this possible. 
Alternatively, independent measurements of the relevant cross sections
away from the signal region could be used to reduce the theoretical
uncertainties. 

If $f_S=f_B=f$, $\log L$ can be
minimized analytically and one finds  the minimum of $\log L$ to occur
at
\begin{equation}
f=\frac{1}{2}\left(
1-(\Delta f)^2N+\sqrt{(1-(\Delta f)^2N)^2+4(\Delta f)^2N_0}
\right) \, ,
\end{equation}
where
\begin{equation}
N=\sum_i(S_i+B_i)
\end{equation}
is the total number of events, 
\begin{equation}
N_0=\sum_i n_{0i}
\end{equation}
the total number of reference events, and $\Delta f$ is the 
uncertainty of the reference cross section. 

The 68.3\% confidence level (CL) limits which can be achieved in
$e^+e^-\to ZHH\to jjb\bar bb\bar b$ are listed in Table~\ref{tab:four}.
\begin{table}
\caption[]{\label{tab:four}
Sensitivities achievable at $68.3\%$ CL for the Higgs boson
self-coupling, $\Delta\lambda_{HHH}$ (see Eq.~(\ref{eq:lam})), in
$e^+e^-\to ZHH\to jjb\bar bb\bar b$ for
$m_H=120$~GeV and 140~GeV, $\sqrt{s}=0.5$~TeV and 1~TeV, and several
choices of integrated luminosities. Results are shown for 4 and $\geq
3$ $b$-tags and, for $\sqrt{s}=0.5$~TeV, two choices of $b$-tagging
efficiencies and 
misidentification probabilities. I assume the uncertainties listed in
Table~\ref{tab:three}. The cuts imposed are described in
Secs.~\ref{sec:three} and~\ref{sec:four}. } 
\vspace{2mm}
\begin{tabular}{c|ccc}
\multicolumn{4}{c}{$\epsilon_b=0.9$, $P_{c\to b}=0.1$,
$P_{j\to b}=0.005$} \\
\hline
\phantom{j} & 0.5~ab$^{-1}$ & 1~ab$^{-1}$ & 2~ab$^{-1}$ \\
\hline
$\sqrt{s}=0.5$~TeV, $m_H=120$~GeV, 4 $b$-tags & $\begin{array}{c} +0.60
\\[-3pt] {-0.56}\end{array}$ &  $\begin{array}{c} +0.44
\\[-3pt] {-0.41}\end{array}$ & $\begin{array}{c} +0.33
\\[-3pt] {-0.30}\end{array}$ \\
$\sqrt{s}=0.5$~TeV, $m_H=120$~GeV, $\geq 3$ $b$-tags & $\begin{array}{c} +1.4
\\[-3pt] {-5.2}\end{array}$ &  $\begin{array}{c} +0.99
\\[-3pt] {-1.12}\end{array}$, $\begin{array}{c} -3.1
\\[-3pt] {-4.8}\end{array}$ & $\begin{array}{c} +0.70
\\[-3pt] {-0.75}\end{array}$, $\begin{array}{c} -3.5
\\[-3pt] {-4.5}\end{array}$ \\
\hline
$\sqrt{s}=0.5$~TeV, $m_H=140$~GeV, 4 $b$-tags & $\begin{array}{c} +2.6
\\[-3pt] {-7.7}\end{array}$ &  $\begin{array}{c} +2.1
\\[-3pt] {-6.8}\end{array}$ & $\begin{array}{c} +1.7
\\[-3pt] {-2.0}\end{array}$, $\begin{array}{c} -3.3
\\[-3pt] {-6.0}\end{array}$ \\
$\sqrt{s}=0.5$~TeV, $m_H=140$~GeV, $\geq 3$ $b$-tags & $\begin{array}{c} +4.8
\\[-3pt] {-8.9}\end{array}$ &  $\begin{array}{c} +3.6
\\[-3pt] {-7.7}\end{array}$ & $\begin{array}{c} +2.7
\\[-3pt] {-6.9}\end{array}$ \\
\hline
$\sqrt{s}=1$~TeV, $m_H=120$~GeV, 4 $b$-tags & $\begin{array}{c} +0.76
\\[-3pt] {-0.60}\end{array}$ &  $\begin{array}{c} +0.53
\\[-3pt] {-0.45}\end{array}$ & $\begin{array}{c} +0.38
\\[-3pt] {-0.33}\end{array}$ \\
$\sqrt{s}=1$~TeV, $m_H=120$~GeV, $\geq 3$ $b$-tags & $\begin{array}{c} +0.65
\\[-3pt] {-0.58}\end{array}$ &  $\begin{array}{c} +0.46
\\[-3pt] {-0.42}\end{array}$ & $\begin{array}{c} +0.33
\\[-3pt] {-0.31}\end{array}$ \\
\hline
$\sqrt{s}=1$~TeV, $m_H=140$~GeV, 4 $b$-tags & $\begin{array}{c} +1.6
\\[-3pt] {-4.2}\end{array}$ &  $\begin{array}{c} +1.1
\\[-3pt] {-1.0}\end{array}$, $\begin{array}{c} -2.4
\\[-3pt] {-3.5}\end{array}$ & $\begin{array}{c} +0.77
\\[-3pt] {-0.71}\end{array}$\\
$\sqrt{s}=1$~TeV, $m_H=140$~GeV, $\geq 3$ $b$-tags & $\begin{array}{c} +1.4
\\[-3pt] {-4.1}\end{array}$ &  $\begin{array}{c} +1.1
\\[-3pt] {-1.1}\end{array}$, $\begin{array}{c} -2.0
\\[-3pt] {-3.5}\end{array}$ & $\begin{array}{c} +0.78
\\[-3pt] {-0.77}\end{array}$ \\
\hline
\multicolumn{4}{c}{$\epsilon_b=0.8$, $P_{c\to b}=0.02$,
$P_{j\to b}=0.001$, $\geq 3$ $b$-tags} \\
\hline
\phantom{j} & 0.5~ab$^{-1}$ & 1~ab$^{-1}$ & 2~ab$^{-1}$ \\
\hline
$\sqrt{s}=0.5$~TeV, $m_H=120$~GeV & $\begin{array}{c} +1.0
\\[-3pt] {-0.8}\end{array}$, $\begin{array}{c} -4.2
\\[-3pt] {-4.9}\end{array}$ &  $\begin{array}{c} +0.80
\\[-3pt] {-0.65}\end{array}$ & $\begin{array}{c} +0.60
\\[-3pt] {-0.51}\end{array}$ \\
\hline
$\sqrt{s}=0.5$~TeV, $m_H=140$~GeV & $\begin{array}{c} +5.1
\\[-3pt] {-8.6}\end{array}$ &  $\begin{array}{c} +3.7
\\[-3pt] {-7.3}\end{array}$ & $\begin{array}{c} +2.7
\\[-3pt] {-6.4}\end{array}$ \\
\end{tabular}
\end{table}
For a Higgs boson with mass close to the current lower mass
bound~\cite{ewk08}, $\lambda_{HHH}$ can be measured with a precision of
$30-60\%$ in $e^+e^-\to jjb\bar bb\bar b$ at $\sqrt{s}=0.5$~TeV with an
integrated luminosity of $0.5-2$~ab$^{-1}$, if one requires
4~$b$-tags. This result is in qualitative agreement with that reported
in Ref.~\cite{tim}. Since the $ZHH$ cross section falls with increasing
center of mass 
energy, the sensitivities which can be achieved for the same final state
and Higgs boson mass at $\sqrt{s}=1$~TeV are slightly worse. While
$\lambda_{HHH}$ can be measured with reasonable precision in $jjb\bar
bb\bar b$
production for $m_H=120$~GeV and $\sqrt{s}=0.5$~TeV, the significantly
reduced signal cross section 
and the increased background make a measurement of the Higgs boson
self-coupling very difficult for $m_H=140$~GeV in this channel. The
sensitivities can be improved somewhat by including other final states
such as $HH\to 4jb\bar b$, however, the resulting limits are still
extremely loose. 

The bounds presented here for $m_H=140$~GeV and $\sqrt{s}=0.5$~TeV are
considerably worse than those derived from the cross section analysis
presented in Ref.~\cite{LC_HH4}. This can be traced to the large
non-resonant background, and the considerable theoretical uncertainty in
the cross section. Both were not taken into account in the earlier
analysis. 

At a 1~TeV machine, the sensitivity limits which can be achieved in the
$jjb\bar bb\bar b$
final state with 4 $b$-tags and $m_H=140$~GeV are approximately a
factor~2 less stringent than those found for $m_H=120$~GeV. Although
the signal cross section increases by about a factor~1.5 including the
final state with 3 tagged $b$-quarks, the much increased
reducible background, combined with the substantial renormalization
uncertainty of the background, ruin the gain from the increased signal
cross section for $\sqrt{s}=0.5$~TeV. For $\sqrt{s}=1$~TeV, 
a slight improvement in the sensitivity limits is observed by
including the final state with three tagged $b$-quarks. 

A straightforward technique to reduce the (reducible) background is to
choose a $b$-tagging efficiency somewhat smaller than that of
Eq.~(\ref{eq:eff1}) for which the charm and light quark/gluon mistagging
probabilities are lower. Table~\ref{tab:four} shows that choosing the
parameters of Eq.~(\ref{eq:eff2}) does indeed improve the
limits which can be achieved for $\sqrt{s}=0.5$~TeV and $\geq 3$
$b$-tags, however, the gain is not sufficient to compensate for the
increased background and theoretical (renormalization scale)
uncertainty which results from including final states with
3~$b$-tags. Nevertheless, a dedicated search for the $b$-tagging efficiency 
and charm and light quark/gluon mistagging probability which optimizes
the sensitivity limits for $\lambda_{HHH}$ may prove beneficial. For a
first step in this direction, see Ref.~\cite{Boumediene:2008hs}. For
$\sqrt{s}=1$~TeV, the background is always relatively small and not much
is gained by varying $\epsilon_b$ and the charm and light quark/gluon
jet misidentification probabilities.

One of the reasons for the weak sensitivity bounds for $\geq 3$ tagged
$b$-quarks is the large renormalization uncertainty of the background
cross section. Reducing the renormalization uncertainty requires either the
calculation of the NLO QCD corrections for the main background sources,
$b\bar bcjjj$, $b\bar b4j$ and $b\bar bc\bar cjj$ production, or a precise
measurement of the cross section of the processes. It is interesting to
investigate 
how much the sensitivity limits would actually improve if the
normalization uncertainty of the background could be reduced to $\Delta
f=10\%$. The sensitivity bounds which one can hope to achieve for
$\sqrt{s}=0.5$~TeV, $\geq 3$ tagged $b$-quarks and the uncertainties
listed in Table~\ref{tab:three}, and those obtained for $\Delta f=10\%$,
are compared in Table~\ref{tab:five}. 
\begin{table}
\caption[]{\label{tab:five}
Sensitivities achievable at $68.3\%$ CL for the Higgs boson
self-coupling, $\Delta\lambda_{HHH}$ (see Eq.~(\ref{eq:lam})), in
$e^+e^-\to ZHH\to jjb\bar bb\bar b$ with $\geq 3$ $b$-tags for
$m_H=120$~GeV and 140~GeV, $\sqrt{s}=0.5$~TeV, and several
choices of integrated luminosities. Results are shown for the
uncertainties listed in 
Table~\ref{tab:three} and for $\Delta f=10\%$, and for two choices of
$b$-tagging efficiencies and 
misidentification probabilities. The cuts imposed are
described in Secs.~\ref{sec:three} and~\ref{sec:four}. } 
\vspace{2mm}
\begin{tabular}{c|ccc}
\multicolumn{4}{c}{$\epsilon_b=0.9$, $P_{c\to b}=0.1$,
$P_{j\to b}=0.005$} \\
\hline
\phantom{j} & 0.5~ab$^{-1}$ & 1~ab$^{-1}$ & 2~ab$^{-1}$ \\
\hline
$m_H=120$~GeV, $\Delta f=100\%$ & 
$\begin{array}{c} +1.4
\\[-3pt] {-5.2}\end{array}$ &  $\begin{array}{c} +0.99
\\[-3pt] {-1.12}\end{array}$, $\begin{array}{c} -3.1
\\[-3pt] {-4.8}\end{array}$ & $\begin{array}{c} +0.70
\\[-3pt] {-0.75}\end{array}$, $\begin{array}{c} -3.5
\\[-3pt] {-4.5}\end{array}$ \\
$m_H=120$~GeV, $\Delta f=10\%$  & $\begin{array}{c} +0.86
\\[-3pt] {-1.04}\end{array}$, $\begin{array}{c} -4.1
\\[-3pt] {-5.1}\end{array}$ & $\begin{array}{c} +0.66
\\[-3pt] {-0.74}\end{array}$, $\begin{array}{c} -4.2
\\[-3pt] {-4.7}\end{array}$ & $\begin{array}{c} +0.52
\\[-3pt] {-0.56}\end{array}$ \\
\hline
$m_H=140$~GeV, $\Delta f=100\%$ & 
$\begin{array}{c} +4.8
\\[-3pt] {-8.9}\end{array}$ &  $\begin{array}{c} +3.6
\\[-3pt] {-7.7}\end{array}$ & $\begin{array}{c} +2.7
\\[-3pt] {-6.9}\end{array}$ \\
$m_H=140$~GeV, $\Delta f=10\%$  & $\begin{array}{c} +3.5
\\[-3pt] {-8.0}\end{array}$ & $\begin{array}{c} +2.8
\\[-3pt] {-7.5}\end{array}$ & $\begin{array}{c} +2.3
\\[-3pt] {-6.8}\end{array}$ \\
\hline
\multicolumn{4}{c}{$\epsilon_b=0.8$, $P_{c\to b}=0.02$,
$P_{j\to b}=0.001$} \\
\hline
\phantom{j} & 0.5~ab$^{-1}$ & 1~ab$^{-1}$ & 2~ab$^{-1}$ \\
\hline
$m_H=120$~GeV, $\Delta f=40\%$  & $\begin{array}{c} +1.0
\\[-3pt] {-0.8}\end{array}$, $\begin{array}{c} -4.2
\\[-3pt] {-4.9}\end{array}$ &  $\begin{array}{c} +0.80
\\[-3pt] {-0.65}\end{array}$ & $\begin{array}{c} +0.60
\\[-3pt] {-0.51}\end{array}$ \\
\hline
$m_H=120$~GeV, $\Delta f=10\%$  & $\begin{array}{c} +0.63
\\[-3pt] {-0.63}\end{array}$ &  $\begin{array}{c} +0.47
\\[-3pt] {-0.46}\end{array}$ & $\begin{array}{c} +0.36
\\[-3pt] {-0.35}\end{array}$ \\
\hline
$m_H=140$~GeV, $\Delta f=100\%$  & $\begin{array}{c} +5.1
\\[-3pt] {-8.6}\end{array}$ &  $\begin{array}{c} +3.7
\\[-3pt] {-7.3}\end{array}$ & $\begin{array}{c} +2.7
\\[-3pt] {-6.4}\end{array}$ \\
\hline
$m_H=140$~GeV, $\Delta f=10\%$  & $\begin{array}{c} +2.5
\\[-3pt] {-7.7}\end{array}$ &  $\begin{array}{c} +1.9
\\[-3pt] {-6.9}\end{array}$ & $\begin{array}{c} +1.5
\\[-3pt] {-6.3}\end{array}$ \\
\end{tabular}
\end{table}
While a reduced theoretical uncertainty will substantially improve the
sensitivity limits for $\geq 3$ tagged $b$-quarks, more stringent limits
than those found for 4 tagged $b$-quarks are only found for
$m_H=140$~GeV and the efficiencies of Eq.~(\ref{eq:eff2}). Of course,
further improvements may be possible by optimizing the $b$-tagging
efficiency and charm and light quark/gluon misidentification
probabilities. 
The sensitivity limits which can be achieved for $\lambda_{HHH}$ in
$e^+e^-\to\nu\bar\nu HH$ with the theoretical uncertainties of
Table~\ref{tab:three} are listed in Table~\ref{tab:six}. For
$m_H=140$~GeV ($m_H=180$~GeV), the combined limits from $e^+e^-\to\nu\bar\nu
b\bar bb\bar b$ and $e^+e^-\to\nu\bar\nu 4jb\bar b$
($e^+e^-\to\nu\bar\nu\ell\nu_\ell 6j$ and $e^+e^-\to\nu\bar\nu 8j$) are
shown. Since the $b\bar bb\bar b$ and $b\bar bjj$ backgrounds do not
affect the $M_{HH}$ differential cross section in the region sensitive
to the Higgs boson self-coupling (see Secs.~\ref{sec:fivea}
and~\ref{sec:fiveb}), they have not been taken into account in the
analysis.
\begin{table}
\caption[]{\label{tab:six}
Sensitivities achievable at $68.3\%$ CL for the Higgs boson
self-coupling, $\Delta\lambda_{HHH}$ (see Eq.~(\ref{eq:lam})), in
$e^+e^-\to\nu\bar\nu HH$ for 
$\sqrt{s}=1$~TeV and 3~TeV and several choices of integrated
luminosities. For $m_H=120$~GeV, results are shown for the
$\nu\bar\nu b\bar bb\bar b$ final state with $\geq 3$ tagged
$b$-quarks. For $m_H=140$~GeV, the combined limits from $e^+e^-\to\nu\bar\nu
b\bar bb\bar b$ with $\geq 3$ $b$-tags and $e^+e^-\to\nu\bar\nu 4jb\bar
b$ are listed, and for $m_H=180$~GeV the combined limits from 
$e^+e^-\to\nu\bar\nu\ell\nu_\ell 6j$ and $e^+e^-\to\nu\bar\nu 8j$ are
given. The cuts imposed are 
described in Secs.~\ref{sec:three} and~\ref{sec:five}. The theoretical
uncertainties used are listed in Table~\ref{tab:three}. } 
\vspace{2mm}
\begin{tabular}{c|ccc}
\multicolumn{4}{c}{$\sqrt{s}=1$~TeV} \\
\hline
\phantom{j} & 0.5~ab$^{-1}$ & 1~ab$^{-1}$ & 2~ab$^{-1}$ \\
\hline
$m_H=120$~GeV & 
$\begin{array}{c} +0.58
\\[-3pt] {-0.27}\end{array}$ &  $\begin{array}{c} +0.30
\\[-3pt] {-0.21}\end{array}$ & $\begin{array}{c} +0.20
\\[-3pt] {-0.17}\end{array}$ \\
\hline
$m_H=140$~GeV & 
$\begin{array}{c} +0.99
\\[-3pt] {-0.41}\end{array}$ &  $\begin{array}{c} +0.94
\\[-3pt] {-0.38}\end{array}$ & $\begin{array}{c} +0.78
\\[-3pt] {-0.25}\end{array}$ \\
\hline
$m_H=180$~GeV & 
$\begin{array}{c} +0.56
\\[-3pt] {-0.32}\end{array}$ &  $\begin{array}{c} +0.55
\\[-3pt] {-0.29}\end{array}$ & $\begin{array}{c} +0.59
\\[-3pt] {-0.28}\end{array}$ \\
\hline
\multicolumn{4}{c}{$\sqrt{s}=3$~TeV} \\
\hline
\phantom{j} & 1~ab$^{-1}$ & 2~ab$^{-1}$ & 3~ab$^{-1}$ \\
\hline
$m_H=120$~GeV & 
$\begin{array}{c} +0.14
\\[-3pt] {-0.12}\end{array}$ &  $\begin{array}{c} +0.11
\\[-3pt] {-0.10}\end{array}$ & $\begin{array}{c} +0.10
\\[-3pt] {-0.09}\end{array}$ \\
\hline
$m_H=140$~GeV & 
$\begin{array}{c} +0.15
\\[-3pt] {-0.19}\end{array}$ &  $\begin{array}{c} +0.15
\\[-3pt] {-0.15}\end{array}$ & $\begin{array}{c} +0.11
\\[-3pt] {-0.14}\end{array}$ \\
\hline
$m_H=180$~GeV & 
$\begin{array}{c} +0.16
\\[-3pt] {-0.20}\end{array}$ &  $\begin{array}{c} +0.15
\\[-3pt] {-0.13}\end{array}$ & $\begin{array}{c} +0.12
\\[-3pt] {-0.12}\end{array}$ \\
\end{tabular}
\end{table}
The bounds on the Higgs self-coupling which can be achieved in
$e^+e^-\to\nu\bar\nu HH$ for $m_H=120$~GeV and 140~GeV at
$\sqrt{s}=1$~TeV are seen to be considerably more stringent than those
found for $e^+e^-\to ZHH$ (see Table~\ref{tab:four}). For
$\sqrt{s}=3$~TeV, $\lambda_{HHH}$ can be measured with a precision of
$10-20\%$ for the Higgs boson masses and the integrated luminosities
considered here. If the theoretical uncertainty on the cross section for
$m_H=180$~GeV can be reduced to 10\%, the bounds listed in
Table~\ref{tab:six} for $\sqrt{s}=3$~TeV improve by a factor
$1.6-2$. For $m_H=140$~GeV ($m_H=180$~GeV), the sensitivity limits  are
expected to improve by roughly a factor~1.1 (1.2) if additional final
states are included in the analysis.

The bounds derived in this Section should be compared with those one
hopes to achieve at the LHC, a luminosity upgraded LHC (SLHC) and a Very
Large Hadron Collider
(VLHC)~\cite{BPR,BPR2}\footnote{Ref.~\cite{BPR,BPR2} quotes 95\% CL
limits. I have recalculated the sensitivity limits for 68.3\% CL with
otherwise unchanged parameters. The 68.3\% CL limits are used for the
comparison presented here.}. For the Higgs mass range
allowed by current experimental
data~\cite{ewk08,Phenomena:2009pt,Goebel:2009qy}, the 
LHC will not be able to probe the Higgs boson self-coupling. At the
SLHC, with an integrated luminosity of 6~ab$^{-1}$, $\lambda_{HHH}$ can
be determined with an accuracy of $50-70\%$ for $m_H=120$~GeV from
$HH\to b\bar b\gamma\gamma$, and $10-15\%$ for $m_H=180$~GeV using the
$HH\to 4W\to\ell^\pm{\ell'}^\pm p\llap/_T 4j$ final state. There are not
enough $HH\to b\bar b\gamma\gamma$ events for $m_H=140$~GeV for a viable
statistical analysis at the SLHC. A luminosity upgrade of the LHC may
be realized at roughly the same time or earlier than an ILC with
$\sqrt{s}=0.5$~TeV. Tables~\ref{tab:three} and~\ref{tab:four} show that,
for $m_H=120$~GeV, the ILC can measure the Higgs boson self-coupling
with considerably higher precision than the SLHC. A measurement of
$\lambda_{HHH}$ for $m_H=140$~GeV will be very difficult at both
machines. For $m_H=180$~GeV, the SLHC shows a clear advantage, however,
a recent combination of electroweak precision data and direct limits
from the Tevatron experiments excludes a SM Higgs boson with
$m_H=180$~GeV at more than $2\sigma$~\cite{Goebel:2009qy}.

An upgraded ILC with $\sqrt{s}=1$~TeV considerably improves the chances
to determine $\lambda_{HHH}$ for $m_H=140$~GeV (see
Tables~\ref{tab:four} and~\ref{tab:six}). For $m_H=180$~GeV, however,
the SLHC still promises a more precise measurement. Since sensitivity
bounds scale roughly like $(\int{\cal L}dt)^{(1/4)}$, this statement is
true as long as an integrated luminosity larger than 2~ab$^{-1}$ can be
accumulated at the SLHC.

The results for a 3~TeV $e^+e^-$ collider (CLIC) should be compared with
those which one can hope to achieve at a VLHC. Such a machine, assuming
a center of mass energy of $\sqrt{s}=200$~TeV and an integrated
luminosity of 600~fb$^{-1}$, can determine the Higgs boson
self-coupling for $m_H=120$~GeV ($m_H=140$~GeV) in the $HH\to b\bar
b\gamma\gamma$ final state with a precision of $30-50\%$ ($40-60\%$). At
CLIC, on the other hand, a precision of $10-20\%$ can be achieved for
both values of $m_H$. For $m_H=180$~GeV, a 
VLHC can achieve a precision of about 2\%. Assuming that a 10\% theoretical
uncertainty on the cross section can eventually be realized, the
sensitivity of a 3~TeV $e^+e^-$ collider obtained from the combined
$\nu\bar\nu\ell\nu_\ell 6j$ and $\nu\bar\nu 8j$ final states is limited
to $7-12\%$. The two final states together account for about 43\% of all
$HH\to 4W$ decays. Including more $4W$ final states thus is expected to
improve these limits by roughly a factor~$2^{1/4}\approx 1.2$. A VLHC
therefore may yield more precise information on the Higgs boson
self-coupling than CLIC for $m_H=180$~GeV. 

\section{Summary and Conclusions}
\label{sec:seven}

After discovery of an elementary Higgs boson at the LHC, and
tests of its fermionic and gauge boson couplings, experimental evidence
that the shape of the Higgs potential has the form required for
electroweak symmetry breaking will complete the proof that both, fermion
and weak boson masses, are generated by spontaneous symmetry
breaking. To probe the Higgs potential, one must determine the Higgs
boson self-coupling. 

Only Higgs boson pair production at colliders can accomplish
this. Several years ago, studies have appeared in the
literature~\cite{SLHC,BPR,BPR2,Baur:2003gpa,blondel}, 
exploring the potential of the LHC, a luminosity upgraded LHC, and a
Very Large Hadron Collider to probe the Higgs boson self-coupling. There
are also numerous analyses of Higgs boson pair production at $e^+e^-$
colliders~\cite{LC_HH1a,LC_HH1,LC_HH2a,LC_HH2,LC_HH3,LC_HH4,Yasui:2002se,Boumediene:2008hs,Takubo:2009rn,Giannelli:2009fe}
which usually only explore one particular Higgs mass and/or final state,
or only one center of mass energy. Furthermore, backgrounds are
estimated using a leading-log shower approximation, and the effect on
non-resonant diagrams is not taken into account. 

In this paper, I have presented a more comprehensive analysis of Higgs
pair production in $e^+e^-$ collisions. Both $ZHH$ and $\nu\bar\nu HH$
production have been investigated for several Higgs boson masses and
center of mass energies. The cross section for $\nu\bar\nu HH$
production is much smaller than that for $ZHH$ production for
$\sqrt{s}=0.5$~TeV and below. However, it grows quickly with increasing
energies, and becomes the dominant source of Higgs boson pairs for
$\sqrt{s}\geq 1$~TeV. $ZHH$ ($\nu\bar\nu HH$) production therefore was
studied 
for $\sqrt{s}\leq 1$~TeV ($\sqrt{s}\geq 1$~TeV) only. Acceptance cuts and
minimal detector effects in form of Gaussian smearing, as well as the
energy loss of $b$-quarks were taken into account in all
calculations. The efficiencies and misidentification probabilities used
are summarized in Eqs.~(\ref{eq:eff1}) and~(\ref{eq:eff2}). The
distribution of the Higgs
pair invariant mass, $M_{HH}$, was found to be sensitive to the
Higgs boson self-couplings and was used for a log-likelihood based
sensitivity analysis. 

For $ZHH$ production, I concentrated on the $ZHH\to jjb\bar bb\bar b$
final state, requiring that the two light quark jets are consistent with a
$Z$-boson. Requiring four tagged $b$-quarks, the $jjb\bar bc\bar c$ and
$b\bar b4j$ backgrounds were found to be relatively small for
$m_H=120$~GeV and $\sqrt{s}=0.5$~TeV. Non-resonant diagrams change the
$jjb\bar bb\bar b$ cross section by about 10\%. For $m_H=140$~GeV, on
the other 
hand, taking into account the non-resonant diagrams roughly doubles the
cross section, while the $jjb\bar bc\bar c$ background can easily be as
large as the signal cross section. This, together with the extremely
small signal cross section, and the large renormalization uncertainty of
the reducible background cross section, implies that it will be extremely
difficult to measure the Higgs boson self-coupling in $jjb\bar bb\bar
b$ production 
with four tagged $b$-quarks and $m_H=140$~GeV at an ILC with a center of
mass energy of 0.5~TeV. At $\sqrt{s}=1$~TeV, the reducible and
non-resonant backgrounds are significantly smaller than at a 0.5~TeV
collider, and a rough measurement of $\lambda_{HHH}$ in $e^+e^-\to
jjb\bar bb\bar b$ may be possible for $m_H=140$~GeV. 

To increase the signal cross section, one can include other $Z$ and/or
Higgs decay final states. Since $B(Z\to\ell^+\ell^-)$ is small, there
will be little gain by including the $\ell^+\ell^- b\bar bb\bar b$ final
state in 
the analysis. For $m_H=140$~GeV, $B(H\to WW^*\to 4j)\approx 23\%$. If
one of the Higgs bosons in $ZHH$ production decays into four jets, the
final state 
consists of $b\bar b6j$. Since the combinatorial background complicates
identification of which jets originate from Higgs and which from
$Z$-boson decays, I have not considered $b\bar b6j$ production here.

A more straightforward approach is to increase the signal cross section
by taking into account final states where only three $b$-quarks
are tagged. Requiring only three tagged $b$-quarks strongly increases
the reducible background and this more than compensates the potential gain
in sensitivity from the increased 
signal cross section. However, the reducible background can be reduced
by choosing a different working point in $\epsilon_b - P_{c\to b} -
P_{j\to b}$ space. Replacing the efficiencies and misidentification
probabilities of Eq.~(\ref{eq:eff1}) with those of Eq.~(\ref{eq:eff2}),
however, yields only a minor improvement in the sensitivity to
$\lambda$. Nevertheless, a search for an 
optimal working point similar to that in Ref.~\cite{Boumediene:2008hs}
may be promising.

For $\sqrt{s}=1$~TeV, the $\nu\bar\nu b\bar bb\bar b$ cross section is
about a factor two larger than that for $jjb\bar bb\bar b$
production. Furthermore, the 
reducible background is small, both for $m_H=120$~GeV and
140~GeV. Non-resonant diagrams modify the $M_{HH}$ differential cross
section by ${\cal O}(10\%)$
except close to the $HH$ threshold where they have a much larger effect.
The effect from non-resonant diagrams therefore can be easily confused
with a positive anomalous Higgs self-coupling, $\lambda_{HHH}>0$ (see
Figs.~\ref{fig:six}a and~\ref{fig:seven}a). Due to the larger cross
section, and the reduced background, the bounds which can be obtained
for $m_H=120$~GeV are significantly better than those one may hope to
achieve in $ZHH\to jjb\bar bb\bar b$. 

For $m_H=140$~GeV, the signal cross section can be increased by including
the $\nu\bar\nu b\bar b4j$ final state which results when one of the
Higgs bosons decays via $H\to WW^*\to 4j$. Combining the limits from
$\nu\bar\nu b\bar bb\bar b$ and $\nu\bar\nu b\bar b4j$ production, I
found that the 
Higgs boson self-coupling can be probed with $25-80\%$ accuracy with
2~ab$^{-1}$ at a 1~TeV $e^+e^-$ collider. Such a machine will also be
able to probe $\lambda$ for a heavier Higgs boson. For $m_H=180$~GeV,
and utilizing the two final states with the largest branching ratios,
$HH\to 4W\to\ell^\pm\nu_\ell 6j$ and $HH\to 4W\to 8j$, I found that a
1~TeV $e^+e^-$ collider can probe $\lambda$ with an accuracy of
$30-60\%$. A 3~TeV collider will be able to measure the Higgs boson
self-coupling with a precision of $10-20\%$ for the Higgs boson
masses investigated here. 

A {\tt MadEvent} based calculation of the full set of non-resonant
Feynman diagrams contributing to $\nu\bar\nu b\bar 4j$,
$\nu\bar\nu\ell\nu_\ell 6j$ and $\nu\bar\nu 8j$ production is currently
not feasible as it requires very large computing resources. Other
programs, which are not based on the evaluation of Feynman diagrams
promise a solution for this problem in the future, however, these
programs are currently not able to handle the final states of
interest. In absence of a calculation which includes the full set of
non-resonant Feynman diagrams, I have estimated their effect from
calculations which include sub-sets of the non-resonant Feynman
diagrams. Unavoidably, this introduces a theoretical uncertainty in the
cross section which I have assumed to be of the same size as the effect
from those non-resonant diagrams which were included in the
calculation. If this uncertainty can be reduced in the future by a
calculation which includes the full set of non-resonant Feynman
diagrams, the bounds which can be achieved will improve by a
factor~1.6~--~2. 

Uncertainties in the calculations presented in this paper also originate
from the detector performance assumed. A better resolution of the
hadronic calorimeter than that assumed such as that promised by the
particle flow algorithm~\cite{Magill:2007zz} may make it possible to
tighten the invariant mass 
window for $b$-jet and light quark/gluon jet pairs. This would reduce
both reducible and irreducible backgrounds and thus improve the
sensitivity to anomalous Higgs boson self-couplings. The reducible
backgrounds in final states involving $b$-quarks strongly depend on the
charm and light quark/gluon jet misidentification probabilities. More
powerful algorithms for tagging $b$-quarks may well reduce the
misidentification probabilities in the future. 

One of the main results of the calculations presented here is that a
0.5~TeV ILC will be able to determine the Higgs boson self-coupling only
if the Higgs boson mass is rather close to the current lower bound. In
contrast, at a 1~TeV $e^+e^-$ linear collider it will be possible to
measure $\lambda$ for larger values of $m_H$, and to reach a
significantly higher precision than for $\sqrt{s}=0.5$~TeV. The
precision which can be reached at a 1~TeV ILC for $m_H\leq 140$~GeV is
considerably better than that one can hope to achieve at the SLHC or a
VLHC. For a heavier Higgs boson with $m_H=180$~GeV, which is disfavored
by a global fit to electroweak data and current direct limits, on the
other hand, 
the SLHC and VLHC promise better limits. Even a 3~TeV $e^+e^-$ collider
will not be able to measure the Higgs boson self-coupling for
$m_H=180$~GeV more accurately than a VLHC. 

\acknowledgements
I would like to thank J.~Reuter for his help with {\tt WHIZARD} and for
useful discussions, and F.~Krauss for an in-depth discussion on the
capabilities of {\tt Sherpa}. I would also like to thank the High Energy
Physics 
Group at Michigan State University, where most of this work was done,
for their generous hospitality. This research was supported by the
National Science Foundation under grants No.~PHY-0456681 and PHY-0757691.


\bibliographystyle{plain}

\end{document}